\documentclass[aps,prb,groupedaddress,floats,reprint]{revtex4-2}

\usepackage[english]{babel}

\usepackage{amsfonts,graphicx,soul,natbib,mathrsfs}

\usepackage[colorlinks=true]{hyperref}

\usepackage{epstopdf}
\usepackage{textcomp}
\usepackage{xcolor}
\usepackage{soul}
\usepackage{physics}
\usepackage{graphics}
\usepackage{comment}
\usepackage{amssymb}

\renewcommand{\vec}[1]{\textnormal{\boldmath$#1$}}
\newcommand{\expv}[1]{\left\langle #1 \right\rangle}

\usepackage{dcolumn}
\usepackage{bm}
\begin{document}

\title{Evolution of the accelerated charged vortex particle \\ in an inhomogeneous magnetic lens.}

\author{S.S. Baturin}%
\email{s.s.baturin@gmail.com}%
\affiliation{School of Physics and Engineering,
ITMO University, 197101 St. Petersburg, Russia}%

\author{D.V. Grosman}%
\email{dmitriy.grosman@metalab.ifmo.ru}%
\affiliation{School of Physics and Engineering,
ITMO University, 197101 St. Petersburg, Russia}%

\author{G.K. Sizykh}%
\affiliation{School of Physics and Engineering,
ITMO University, 197101 St. Petersburg, Russia}%

\author{D.V. Karlovets}%
\affiliation{School of Physics and Engineering,
ITMO University, 197101 St. Petersburg, Russia}%

\date{\today}
\begin{abstract}
   We present a detailed analysis of the capture and acceleration of a non-relativistic charged vortex particle (electron, positron, proton, etc.)  with an orbital angular momentum in a field of an axisymmetric electromagnetic lens, typical for a linear accelerator. We account for the acceleration as well as for the inhomogeneity of both electric and magnetic fields that may arise from some real-life imperfections. We establish conditions when the wave packet can be captured and successfully transported through the lens. We describe the transition process and explain how a free Laguerre-Gaussian packet could be captured into the Landau state of the lens preserving its structure for all moments in time. Several representative examples are provided to illustrate developed formalism.  
\end{abstract}

\maketitle

%================================================================================================
\section{Introduction}
%================================================================================================
Waves carrying intrinsic orbital angular momentum (OAM) are widely used both in optics \cite{Allen1999,FrankeArnold2008} and in electron microscopy \cite{Bliokh2007,McMorran2011,Tamburini2006}. The azimuthal gradient of the helical phase produces a spiraling current and a well-defined OAM along the beam axis. 

Over the past 25 years, studies of the vortex states have turned into an extensive field. Photons with a large OAM are generated now not only in the optical but also in the radio and X-ray ranges. Detail could be found in numerous reviews, see for instance Refs.\cite{ct1,Mono,ct3}. Moreover, it turned out that other elementary particles in a state with non-zero projections of the orbital angular momentum can find useful applications in atomic and nuclear physics and even in elementary particle physics and microscopy. Not long ago, the first results on the generation of electrons with a large OAM were achieved, and now there are many theoretical and experimental findings on this topic \cite{ct4}. Several fundamental processes involving electrons with a large OAM are studied in \cite{ct5,ct6,ct7}. Recently, the first experimental work on the generation of vortex cold neutrons was carried out \cite{ct8} as well. The latest achievement was obtained by a group from Israel \cite{ct9} where a method for obtaining vortex atoms and molecules was demonstrated.

Despite significant progress in theory and experimental methods in the field of particles with a large OAM, there are several opened questions and challenges. Namely: production of photons with a large OAM and high coherence in the hard X-ray range and generation of relativistic electrons and protons with a large OAM. The development in this direction is related to the analysis of the possibility of creating sources of high-energy vortex particles based on existing accelerator complexes and the development of an appropriate theoretical and experimental base.

An important step for further development of this direction and the application of such particles to the studies of matter on subnanometer scales (atomic and nuclear physics, high-energy physics) is the creation of sources of vortex high-energy particles for the possibility of irradiation of nano-dimensional structures with X-ray twisted photons, problems of studying nano- and metamaterials with specified properties and fundamental issues of interaction of particles with non-zero OAM and finite spatial coherence with the matter. The generation of vortex electrons with high OAMs and their acceleration to relativistic energies at existing accelerator facilities is of particular interest for particle physics and hadron physics. The infrastructure and the main expensive components for achieving high energies (hundreds or even thousands of MeV) are already available, as they are used exclusively to generate beams without angular momentum. 

However, the dynamics of charged particles with the phase vortices in a lattice that consists of a set of drifts and magnetic lenses have not been investigated previously in detail. The problem was studied excessively in the paraxial approximation for the homogeneous \cite{FarEff,Bliokh2012,Silenko2021,Karlovets2021,JAGA95} as well as for the case of inhomogeneous fields \cite{JAGA89,JAGA90,NUF21}. In the present study, we show how \textit{the paraxial approximation can be avoided} at least in a non-relativistic case and highlight that evolution of the time-dependent wave function is fully described by the evolution of three parameters only: namely, the dispersion, the Gouy phase, and the radius of curvature of the wavefront. We focus on the evolution of the dispersion only and consider time-dependent capture of the free vortex wavepacket by the electromagnetic lens and establish capture and transport conditions in terms of the dispersion that guarantee successful asymptotic transition of the free Laguerre-Gaussian packet to the stationary Landau state.

The dynamics of particles in accelerators is mostly classical in a sense that each particle behaves like a point charge exposed to electromagnetic fields of an accelerating radio frequency (RF) wave, of the electric and magnetic lenses, as well as of the other particles in a bunch. The quantum effects usually arise due to recoil in the emission of photons and in broadening of the classical orbits in storage rings \cite{ST}.
   There is, however, another family of quantum effects that has not been explored yet, neither theoretically nor experimentally: namely, the role of the wave packet shape and topology. While the particle wave packets are described as plane waves in the overwhelming majority of high-energy experiments \cite{BLP, Peskin}, the experimenters at electron microscopes are now able to generate quantum states of 300-keV electrons that cannot be described as plane waves or even as Gaussian packets \cite{McMorran2011}. Such states are called \textit{the vortex electrons} \cite{Bliokh2007} and they represent massive counterparts of the better known twisted photons \cite{Mono, UFN} with the phase vortices and the doughnut-shape of the energy density. The rich physics that can result from further developments in this direction and from bringing the vortex particles into the high-energy domain is outlined in a recent review by I. Ivanov \cite{IvanovPubl}.

If accelerated to relativistic energies of at least tens of MeV, vortex particles (electrons, for instance) can open a new vista in particle physics experiments thanks to different cross sections already at the tree level, to sensitivity to the general phase of the scattering amplitudes, and the packet spatial coherence and shape. Indeed, unlike the plane-wave states with a definite helicity and momentum, $|{\bm p}, \lambda\rangle$, the vortex states are \textit{cylindrical waves} $|p_{\perp}, p_z, j_z, \lambda \rangle$ with a definite projection of the total angular momentum 
$$
\hat{J}_z |p_{\perp}, p_z, j_z, \lambda \rangle = (\ell + \lambda) |p_{\perp}, p_z, j_z, \lambda \rangle
$$ 
onto the propagation axis, but with an undefined azimuthal angle of the momentum,
so that the energy is the same, $\varepsilon = \sqrt{{\bm p}^2 + m^2} = \sqrt{p_{\perp}^2 + p_z^2 + m^2}$. Consequently, the corresponding cross sections involve the conservation law of the total angular momentum of all particles in a reaction, $\delta_{j_z^{\text{in}},j_z^{\text{out}}}$, something the customary cross sections with the plane-wave states are insensitive to. As a result, the vortex electrons and other vortex particles can represent a new effective tool for probing the angular momenta of partons inside a hadron.

Using a linear accelerator (linac) seems to be the best way to accelerate vortex electrons because the angular momentum is conserved in axially symmetric electric and magnetic fields \cite{Karlovets2021}. To get reliable estimates of the needed parameters, there are still many problems to be solved, the first of them is developing a theoretical model of quantum dynamics of the vortex particles in realistic fields of the accelerating RF waves and of the focusing electromagnetic lenses. Spatial inhomogeneity of the accelerating and focusing fields inside a linear accelerator is of high importance for classical dynamics of bunches containing billions of particles with the bunch width varying from a fraction of a millimeter to centimeters \cite{PDG, Reiser}. For quantum dynamics of single-particle wave packets these inhomogeneities, on the contrary, are supposed to play \textit{much smaller role} because the typical width of an electron packet does not exceed \textit{a few nanometers} nearby a cathode \cite{Cho, Cho2013, Lat, Ehberger}, and so a linear approximation (weak inhomogeneity) preserving the packet emittance seems to suffice. 

Indeed, let the electron packet have an RMS radius (transverse coherence length) $\sqrt{\sigma^2_{\perp}(0)}$ of the order of 1 nm. The spatial inhomogeneity of the field can come into play as soon as the packet spreads to the typical width of a multi-particle bunch, say, of 1 mm. The distance $\langle z\rangle$ needed for that can be derived from the following relation \cite{Karlovets2021}
\begin{align}
\frac{\sigma^2_{\perp}(\langle \hat z\rangle)}{\sigma^2_{\perp}(0)} = 1 + \frac{\langle \hat z\rangle^2}{z_R^2},
\end{align}
where the Rayleigh length of an electron packet moving with a mean velocity $\langle u\rangle$ is
\begin{align}
z_R = \langle \hat u\rangle t_c \frac{\sigma^2_{\perp}(0)}{\lambda_c^2}.
\end{align}
Here, $\lambda_c = \hbar/mc \sim 10^{-11}$ cm, $t_c = \lambda_c/c \sim 10^{-21}$ s. For fast electrons, we get
$$
\langle \hat z\rangle \sim 10\, m.
$$
Thus, the drift section much shorter than that guarantee the weakness of the non-linear effects that can potentially alter the packet emittance and destroy the quantum state. Finally, inside an accelerating cavity or in an electromagnetic lens where charge is moving in a static uniform magnetic field the packet RMS radius does not grow but \textit{oscillates} around some value defined by the stationary Landau orbit \cite{Karlovets2021,Silenko2021,FarEff}, very similar to the betatron oscillations in a synchrotron \cite{ST}.

The analysis is organized as follows. First, we derive the equations of motion for the packet RMS-radius $\langle \hat \rho^2 \rangle$ inhomogeneous fields and establish conditions that guarantee the capture and transport of the incoming free packet by the magnetic field. Next, we develop a perturbation theory and calculate first order correction to the $\langle \hat\rho^2 \rangle$ that arises from the field inhomogeneity.

We consider the initial particle to be an electron for definiteness and simplicity with the charge $q = - e$, where $e>0$ is the elementary charge. The result could be easily generalized to other particles with the proper substitution of mass and charge. Throughout the paper, we use the natural system of units: $\hbar=c=1$.

%================================================================================================
\section{LG packets in free space and in magnetic field: Exact wave functions \label{sec:2}}
%================================================================================================
We start with the non-stationary Schr\"{o}dinger's equation
\begin{align}
\label{eq:sch}
    i\frac{\partial \psi}{\partial t} = \hat{H}\psi,
\end{align}
which has \textit{an exact non-paraxial solution} in the form of an LG packet in free space with the following the Hamiltonian
\begin{align}
\label{eq:freeH}
    \hat H_{\mathrm{free}}=\frac{\bm{\hat p}^2}{2m},
\end{align}
where the momentum operator $\bm{\hat p}=-i\nabla$ and 
\begin{align}
\label{eq:mop}
    \hat{\bm{p}}^2 = -\frac{\partial^2}{\partial \rho^2} - \frac{1}{\rho} \frac{\partial}{\partial \rho} + \frac{1}{\rho^2} \hat{L}_z^2 - \frac{\partial^2}{\partial z^2}.
\end{align}
Here we use cylindrical coordinates $(\rho,\phi,z)$, assume the wave packet propagation along the $z$-axis $\langle \vec p \rangle=\left\{0,0,\langle p \rangle \right\}$, and the angular momentum $z$-projection operator is
$$
\hat L_z\equiv-i \frac{\partial}{\partial \phi}.
$$
For the free Hamiltonian, the variables can be separated, consequently, the wave function can be factorized,
\begin{align}
    \psi(\rho,\phi,z,t)&=\psi_\perp(\rho,\phi,t)\psi_\parallel(z,t).
\end{align}
The longitudinal part of the wave function can be expressed via the spatially delocalized plane-wave
\begin{align}
    \psi_\parallel(z,t)\propto\exp(i p_z z - i \frac{p_z^2}{2m}t),
    \label{eq:zpwfree}
\end{align}
with $p_z$ being an eigenvalue of $\hat p_z=-i\partial_z$. 
The transverse part of the wave function can be written as a so-called standard Laguerre-Gaussian (LG) packet that exactly satisfies the Schr\"{o}dinger's equation and reads \cite{PRA19, Karlovets2021} 
\begin{align}
\label{eq:LG}
    &\psi_\perp(\rho,\phi,t)\propto \frac{1}{\sigma_{\perp}(t)}\left(\frac{\rho}{\sigma_{\perp}(t)} \right)^{|l|}\mathcal{L}_{n}^{|l|}\left[\frac{\rho^2}{\sigma^2_{\perp}(t)}\right]\times \\&\exp\left[-\frac{\rho^2}{2\sigma^2_\perp(t)} \right] \exp\left[i l \phi - i\phi_G(t)+i\frac{\rho^2}{2 R^2(t)}\right] \nonumber.  
    %&\int d^2\rho|\psi_{n,l}(\bm{\rho},t)|^2=1, \;
\end{align}
Here $\mathcal{L}_{n}^{|l|}$ is the generalized Laguerre polynomial, $n=0,1,2,...$ is the radial quantum number, $l=0,\pm 1,\pm 2,...$ is the OAM, that is, an eigenvalue of the $\hat L_z$ operator, whereas the functions $\sigma_\perp(t)$, the Gouy phase $\phi_G(t)$ and $R(t)$, which is an equivalent of the wave front curvature, are given by 
\begin{align}
    &\sigma_{\perp}(t)=\frac{1}{\sigma_p}\sqrt{1+\frac{t^2}{t_d^2}},\\
    &\phi_G(t)=(2n+|l|+1) \arctan\left(\frac{t}{t_d}\right), \\
    &R^2(t)=\frac{\sigma^2_{\perp}(t)}{t/t_d}.
\end{align}
Here and above 
$$
t_d= m\sigma_p^{-2} = t_c\, \frac{\sigma_{\perp}^2(0)}{\lambda_c^2}
$$ 
is the diffraction time, $\sigma_p$ is the transverse momentum dispersion at the focal point $t=t_0=0$, and $t_c = \lambda_c/c, \lambda_c = \hbar/mc$.

\begin{figure}[t]
\center{\includegraphics[width=1.\linewidth]{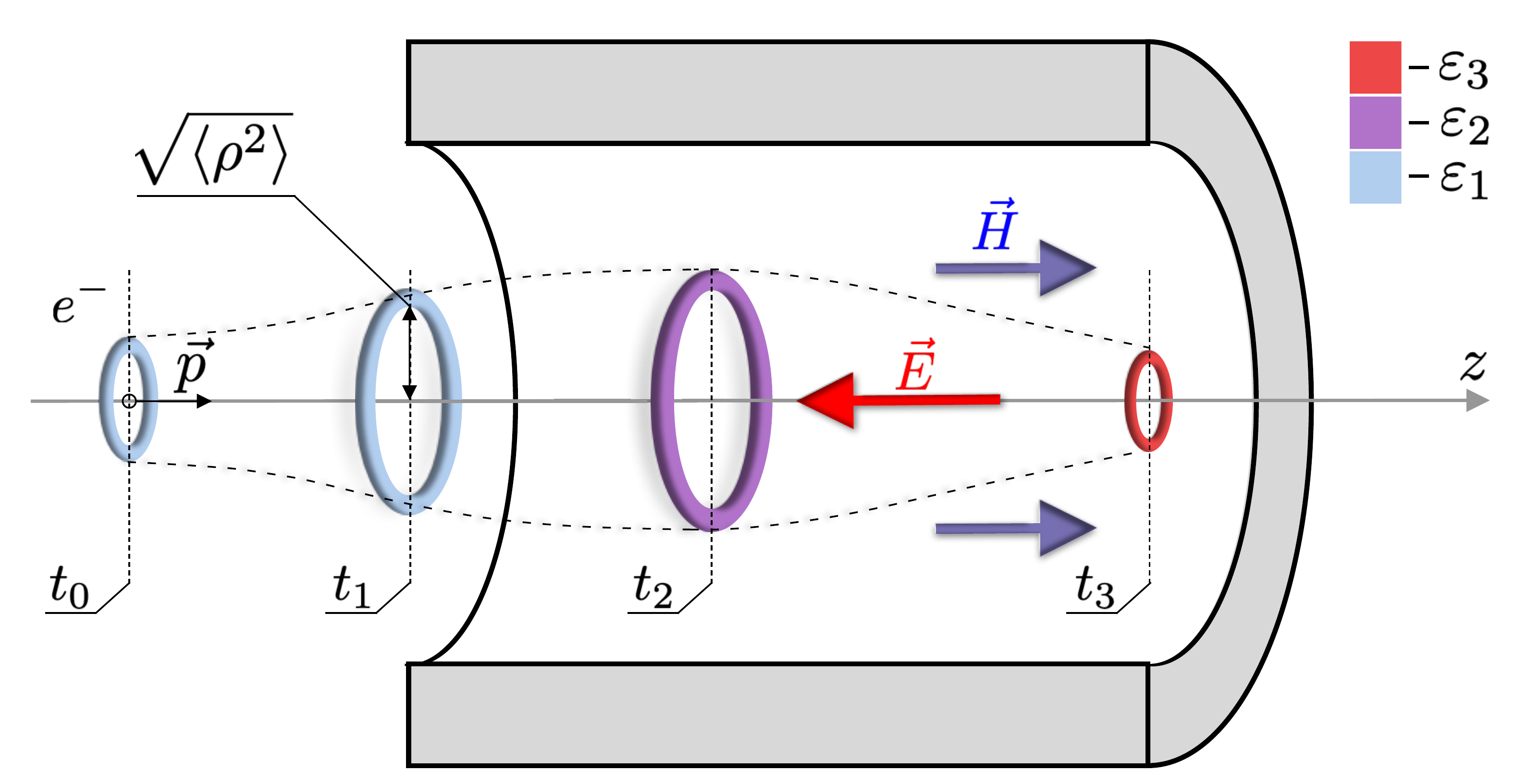}}
\caption{Schematic diagram of the problem. A free LG-electron packet with the total energy $\varepsilon_1$ starts expanding at the time $t=t_0$, next it enters the lens at the time $t=t_1$ and is accelerated to the energy $\varepsilon_3$ and focused while propagating inside the lens.}
\label{fig:lens}
\end{figure}

We consider a free vortex electron described as the LG packet at the moment of time $t=t_1$ that enters an \textit{accelerating cavity} enclosed into a focusing magnetic lens (a solenoid), as shown in Fig.\ref{fig:lens}. We neglect the slippage effects assuming perfect synchronization between the electron and the accelerating RF-wave, and we also neglect edge effects at the entrance of the cavity. The simplest model employs time-independent and spatially homogeneous accelerating electric $E_0$ and magnetic $H_0$, both directed along the $z$ axis.

Now we take into account that the magnetic field $\vec{\mathrm{H}}$ of a real focusing lens as well as the accelerating electric field $\vec{\mathrm{E}}$ \textit{deviate slightly} from the uniform values $H_0$ and $E_0$, and their Tailor decomposition with respect to the radius $\rho$ near the axis of the lens and with respect to $z$ near the entrance of the lens $z=0$ reads \cite{Reiser}:
\begin{align}
\label{eq:elf}
    E_\rho&=-\frac{E_1}{2} \rho +\mathcal{O}\left( \rho^2, \rho z, z^2 \right), \nonumber \\
    E_z&=E_0+E_1 z+\mathcal{O}\left( \rho^2, \rho z, z^2 \right).
\end{align}
for the electric field, and similarly for the magnetic field
\begin{align}
\label{eq:magf}
    H_\rho&=-\frac{H_1}{2} \rho +\mathcal{O}\left( \rho^2, \rho z, z^2 \right), \nonumber \\
    H_z&=H_0+H_1 z+\mathcal{O}\left( \rho^2, \rho z, z^2 \right).
\end{align}

We have introduced the following notations: %\hl{Seems to better define them through z-derivatives or add 2 in denominator at lest}
\begin{align}
    E_1\equiv -2\left. \frac{\partial E_{\rho}}{\partial \rho} \right|_{\rho=0},~~~
    H_1\equiv -2\left. \frac{\partial H_{\rho}}{\partial \rho} \right|_{\rho=0}.
\end{align}
We assume no external currents and it is straightforward to check that the above fields satisfy Maxwell's equations, see Appendix \ref{app:1}.

The corresponding scalar and vector potentials are
\begin{align}
\label{eq:pots}
    &\varphi =  \frac{E_1 \rho^2}{4}- E_0 z - \frac{E_1}{2} z^2, \nonumber \\\; &\bm{\mathrm{A}}=\frac{H_0+H_1 z}{2}\rho\ \bm{e_\phi}
    . 
\end{align}
\\

At the moment $t_1$ the LG packet enters the lens with
the following Hamiltonian:
\begin{align}
\label{eq:inH}
    \hat H_\mathrm{lens}=\frac{\left(\bm{\hat p}^{\mathrm{kin}}\right)^2}{2m}-e\hat \varphi.
\end{align}
Here, for the electron with the charge $q=-e$ the kinetic momentum operator $\bm{\hat p}^{\mathrm{kin}}=\bm{\hat p} + e\bm{\mathrm{A}}=-i \nabla + e\bm{\mathrm{A}}$. With Eq.\eqref{eq:pots} we may rewrite the Hamiltonian Eq.\eqref{eq:inH} as
\begin{align}
\label{eq:lensHam}
    & \hat{H}_\mathrm{lens} = \frac{\hat{\bm{p}}^2}{2m} + \frac{\omega_{c}(\hat z) \hat{L}_z}{2}  + \frac{m}{8} \omega_{c}^2(\hat z) \hat \rho^2 - e\hat \varphi,
\end{align}
with $\omega_c$ being the cyclotron frequency
\begin{align}
    \quad \omega_{c}(\hat z) = \frac{e H_0}{m} + \frac{e H_1}{m} \hat z=\omega_0\left[1 + \kappa_M \frac{\hat z}{L}\right],
\end{align}
with 
\begin{align}
\label{eq:om0}
\omega_0\equiv \frac{eH_0}{m}
\end{align}
and
\begin{align}
\label{eq:aM}
\kappa_M \equiv \frac{L H_1}{H_0},
\end{align}
where $L$ is the length of the lens.  
% and square of the canonical momentum operator is given by Eq.\eqref{eq:mop}. 
Without loss of generality, we can assume $H_0>0$. The case with the opposite sign of $H_0$ could be accounted for by a simple change in the sign of the OAM, $l \to -l$.   

Importantly, due to the axial symmetry of the problem, both the Hamiltonians $\hat{H}_{\mathrm{free}}$ and $\hat{H}_{\mathrm{lens}}$ commute with $\hat{L}_z$, consequently, the electron intrinsic orbital angular momentum \textit{is conserved} during the transport and acceleration inside the cavity.     

For homogeneous fields with $E_1=H_1=0$, there exists an exact solution to the non-stationary Schr\"{o}dinger's equation Eq.\eqref{eq:sch} with $\hat{H}=\hat{H}_\mathrm{lens}$. Indeed, one can split the Hamiltonian $\hat{H}_\mathrm{lens}$ as follows  
\begin{align}
    \hat{H}_\mathrm{lens}=\hat{H}_\perp+\hat{H}_\parallel
\end{align}
with
\begin{align}
\label{eq:prLH}
    \hat{H}_\perp=&- \frac{1}{2m}\left[\frac{1}{\rho}\frac{\partial}{\partial \rho}\left(\rho \frac{\partial}{\partial \rho} \right)+ \frac{1}{\rho^2} \frac{\partial^2}{\partial \phi^2}\right]-\nonumber \\&-i \frac{\omega_0}{2}\frac{\partial}{\partial \phi}  + \frac{m\omega_0^2}{8} \rho^2,
\end{align}
and 
\begin{align}
\label{eq:lnLH}
    \hat{H}_\parallel=-\frac{1}{2m}\frac{\partial^2}{\partial z^2}-e |E_0| z.
\end{align}
Note that we choose $E_0$ such that the particle is accelerated, so regardless of the sign of the charge Eq.\eqref{eq:lnLH} stays valid both for particles carrying positive and negative charge. The longitudinal Hamiltonian and the transverse one commute $[\hat H_\perp,\hat H_\parallel]=0$, consequently, the wave function can be factorized as
\begin{align}
     \Psi(\rho,\phi,z,t)= \Psi_\perp(\rho,\phi,t) \Psi_\parallel(z,t).
\end{align}
The transverse part $\Psi_\perp(\rho,\phi,t)$ is a well known Landau state \cite{Landau,Bliokh2012}
\begin{align}
    \Psi_\perp(\rho,\phi,t) \propto \left(\frac{\rho}{\rho_H}\right)^{|l|}&\mathcal{L}_{n'}^{|l|}\left[\frac{2\rho^2}{\rho_H^2}\right] \times \nonumber \\&\exp\left[-\frac{\rho^2}{\rho_H^2} + i l \phi-i \varepsilon_\perp t\right],
    \label{eq:Landau}
\end{align} 
where $n'=0,1,2,...$ is the radial quantum number, 
\begin{align}
\rho_H=\sqrt{\frac{4}{eH_0}}=\sqrt{\frac{4}{m \omega_0}}
\label{eq:rhom}
\end{align}
is the characteristic radius of the orbit and 
\begin{align}
    \varepsilon_{\perp}=\frac{\omega_0}{2}\left(2n'+|l|+l + 1 \right) 
    \label{eq:LandauE}
\end{align}
is the transverse energy. Note that in contrast to the spreading free LG packet (\ref{eq:LG}) the Landau state does not spread and its transverse "size" (packet dispersion) is now time independent. 

The longitudinal part of the wave function $\Psi_\parallel(z,t)$ for $t>t_1$ can be found by directly solving the equation $i \partial_t \Psi_\parallel(z,t) = \hat H_\parallel \Psi_\parallel(z,t)$ and the result  
\begin{align}
    &\Psi_\parallel(z,t)\propto \\ \nonumber &\exp \left [ip_z(t) z-i\varepsilon_\parallel(t) (t - t_1)-i\frac{p_0^2}{2m}t_1 \right]
\end{align}
with
\begin{align}
        &p_z(t)=p_0+e|E_0|(t-t_1),
\end{align}
and
\begin{align}
\label{eq:zpwln}
    \varepsilon_\parallel(t)=\frac{p_0^2}{2m}& +
    \frac{e|E_0|p_0}{2m}(t-t_1)\nonumber \\&+\frac{e^2 E_0^2}{m}\frac{(t-t_1)^2}{6}
\end{align}
describes \textit{longitudinal acceleration} of the non-relativistic particle. Here, $p_0$ is the longitudinal momentum at the lens entrance ($t=t_1$ in Fig.\ref{fig:lens}).

Note that although longitudinal parts of the wave functions for free space Eq.\eqref{eq:zpwfree} and at the lens entrance Eq.\eqref{eq:zpwln} are "matched" and the transverse Landau state Eq.\eqref{eq:Landau} looks very similar to the transverse part of the LG packet Eq.\eqref{eq:LG}, direct capture of the free LG packet into the Landau state is not possible. 
Indeed divergence of the LG packet leads to the non vanishing time derivative of the LG packet dispersion $d \sigma(t)/dt \neq 0$, while for the Landau state this derivative must be zero $d \langle \hat \rho^2 \rangle /dt =0$.

On the other hand, this close match between the free LG state and Landau state provides a useful insight. Intuitively it looks like one can merge the free and Landau state without much disturbance to the transverse structure of the packet such that the free packet asymptotically transforms into the Landau state and the transverse part of the wave function keeps the same ``doughnut'' shape.

Indeed, the transverse part of the free Shr\"odinger equation looks exactly like the paraxial wave equation for light with the only substitution $t \to z$,
\begin{align}
\label{eq:parSchFr}
 \left[\Delta_\perp+\frac{2i}{\lambda_c}\frac{\partial}{\partial t}\right] \Psi_\perp=0.
\end{align}
Inside the lens, the transverse part of the wave function satisfies a similar paraxial wave equation for light that propagates in a medium with quadratic dependence of the refractive index on the radius \cite{N2opt} (with the same substitution $t \to z$): 
\begin{align}
\label{eq:parSch}
 \left[\Delta_\perp+ i m \omega_0\frac{\partial}{\partial \phi}  -\frac{m^2\omega_0^2}{4}\rho^2+2mi\frac{\partial}{\partial t}\right] \Psi_\perp=0.
\end{align}
This connection has previously been pointed out by several authors (see Ref.\cite{FarEff} and Ref.\cite{Silenko2021}), although only within the paraxial approximation.
We emphasize that the paraxial approach has limited applicability and the non-paraxial effects can become important when the spreading is noticeable, $t \gtrsim t_d$ \cite{Karlovets_paraxial1,Karlovets2021}.

We reiterate the difference between the prior art and present consideration that leads to an important consequence. A free LG packet as well as paraxial Landau modes, introduced in Ref.\cite{Silenko2021} (with the proper substitution $z \to t$) gives an exact solution to the transverse part of the nonstationary Schr\"{o}dinger's equation, consequently no approximations are implied. The structure of the paraxial Landau mode, which is an exact solution to the Eq.\eqref{eq:parSch}, is the same as in free space and given by Eq.\eqref{eq:LG} with the different set of optical functions $\sigma(t)$, $\phi_G(t)$ and $R(t)$ that are to be determined.

Given that the angular momentum is conserved, the evolution of the LG packet is fully defined by the evolution of the RMS radius $\langle\hat\rho^2\rangle$, the Gouy phase, and the radius of curvature $R(t)$. We stress that the discussion above unveils an important fact: the free LG packet \textit{can be directly captured and focused} (defocused) by the magnetic solenoid such that \textit{the packet spatial structure is exactly preserved.} In more detail, the RMS radius and its derivative can be continuous at the boundary,
\begin{align}
\label{eq:cont}
\langle \hat\rho^2\rangle_{\text{free}} = \langle\hat\rho^2\rangle_{\text{lens}},\ \frac{d\langle\hat\rho^2\rangle_{\text{free}}}{dt} = \frac{d\langle\hat\rho^2\rangle_{\text{lens}}}{dt},
\end{align}
which also provides the continuation of the packet transverse emittance \cite{Karlovets2021}. In the analysis below, we thus focus on the evolution of the mean square radius as it provides (along with the evolution of the phase shift) full information on the evolution of the LG state.

%================================================================================================
\section{Dynamics of the packet RMS radius}
%================================================================================================
To tackle the non-stationary problem, we start from the Heisenberg's equation that reads
\begin{align}
    \frac{d \hat Q}{d t}=i\left[\hat{H},\hat Q\right] + \frac{\partial \hat Q}{\partial t}.
    \label{eq:Heis}
\end{align}
It allows us to propagate the observable operator $\hat Q$ (and the observable itself based on the Ehrenfest theorem) from the vacuum to the inner part of the lens.

For the sake of convenience, we introduce the square of the transverse velocity operator $\hat{u}^2_{\perp}$ as
\begin{align}
    \begin{aligned}
    &\hat{u}^2_{\perp}=\frac{2 \hat H_\perp}{m}.
    \end{aligned}
    \label{eq:vel}
\end{align}
We note that as far as both in free space and inside the lens the Hamiltonian $\hat H_\perp={\hat p}^2_{\mathrm {kin}}/2m$ where $\hat p_{\mathrm{kin}}$ is the modulus of the kinetic momentum operator, then $\hat{u}$ is just a kinetic velocity operator of the particle. 
Here $\hat H_\perp$ is given by Eq.\eqref{eq:prLH} inside the lens and $\hat H_\perp=\hat{\bm{p}}^2_\perp/2 m$ in free space.
It turns out that it is convenient to raise the order of Heisenberg's equation and consider the following system for $\hat \rho^2$ and $\hat  u_\perp^2$  
\begin{align}
   \left\{\begin{aligned} &\frac{d^2 \hat \rho^2 }{d t^2}=-\left[\hat{H},\left[\hat{H},\hat \rho^2\right]\right], \\
   &\frac{d \hat u_\perp^2 }{d t}=i\left[\hat{H},\hat u_\perp^2\right].
   \end{aligned}\right.
   \label{eq:mainsys}
\end{align}
where the Hamiltonian is given by either Eq.\eqref{eq:freeH} or Eq.\eqref{eq:lensHam}. 

%================================================================================================
\subsection{Homogeneous lens \label{sec:HL}}
%================================================================================================
In free space (for $t\in [0,t_1]$ in Fig.\ref{fig:lens}) the Hamiltonian is given by Eq.\eqref{eq:freeH}. Evaluating the commutator on the right hand side of Eq.\eqref{eq:mainsys} we arrive at
\begin{align}
   \left\{\begin{aligned} &\frac{d^2 \hat \rho^2 }{d t^2}=2\hat u_\perp^2, \\
   &\frac{d \hat u_\perp^2 }{d t}=0.
   \end{aligned}\right.
   \label{eq:mainFree}
\end{align}
Averaging over the system quantum state, we get for $\langle \hat\rho^2 \rangle(t)$ and $\langle \hat p_\perp^2 \rangle(t)$ 
\begin{align}
\begin{aligned}
    &\langle \hat \rho^2 \rangle(t)=\langle \hat \rho^2 \rangle_0+\partial_t\langle{\hat \rho}^2 \rangle_0 t+\langle \hat u_\perp^2 \rangle_0 t^2, \\
    &\langle \hat u^2_\perp \rangle(t)=\langle \hat u_\perp^2 \rangle_0.
    \end{aligned}
    \label{eq:solFree}
\end{align}
Here $\langle \hat \rho^2 \rangle_0$ is the initial mean-square radius of the packet, $\partial_t\langle{\hat \rho}^2 \rangle_0$ is its expansion rate defined by $\langle \bm{\hat \rho} \bm{\hat p}_\perp \rangle(0)$ and $\langle \hat u_\perp^2 \rangle_0$ is defined by the initial mean-square transverse momentum. Assuming that the focal point for the LG packet is at $t=t_0=0$ (meaning $\langle \bm{\hat \rho} \bm{\hat p}_\perp \rangle(0)=0$) we observe quadratic expansion of the packet with time, as has previously been pointed out \cite{Allen1999,Bliokh2012,Silenko2021,Karlovets2021}.
\begin{align}
\begin{aligned}
    &\langle \hat \rho^2 \rangle(t)=\langle \hat \rho^2 \rangle_0+\langle \hat u_\perp^2 \rangle_0 t^2, \\
    &\langle \hat u^2_\perp \rangle(t)=\langle \hat u_\perp^2 \rangle_0.
    \end{aligned}
    \label{eq:solFree2}
\end{align}
For a homogeneous lens, we set $E_1=H_1=0$ in Eq.\eqref{eq:pots}, and consider the Hamiltonian given by Eq.\eqref{eq:lensHam}. 
Evaluating commutators (see Appendix \ref{sec:app2} for details) in Eq.\eqref{eq:mainsys} we arrive at   
\begin{equation}
\left\{\begin{aligned}    &\frac{d^2 \hat \rho^2 }{d t^2} =\frac{2\hat p_\perp^2}{m^2}  - \frac{\omega_0^2}{2} \hat \rho^2 , \\
    &\frac{d \hat{u}_{\perp}^2}{d t} =0.
\end{aligned} \right.
\label{eq:mainsyslens2}
\end{equation}
With the help of the definition Eq.\eqref{eq:vel}, we write the system in a final form
\begin{equation}
\left\{\begin{aligned}    &\frac{d^2 \hat \rho^2 }{d t^2} =2 \hat u^2_\perp  - \frac{2 \omega_0}{m}\hat L_z -\omega_0^2 \hat \rho^2 , \\
    &\frac{d \hat{u}_{\perp}^2}{d t} =0.
\end{aligned} \right.
\label{eq:mainsysFL}
\end{equation}
We note that as far as kinetic energy is conserved, the mean square of the velocity operator remains the same as in free space, however the mean square radius now oscillates, in full agreement with Refs.\cite{Karlovets2021,Silenko2021,FarEff}. Using the solution in free space at the lens entrance ($t=t_1$) as initial conditions for the system Eq.\eqref{eq:mainsysFL}, we arrive at the solution inside the lens,
\begin{align}
\begin{aligned}
     \langle \hat\rho^2 \rangle (t) &= \langle \hat\rho^2 \rangle_{st} + \left( \langle \hat\rho^2 \rangle_{in} - \langle \hat\rho^2 \rangle_{st} \right) \cos\left[\omega_0 (t-t_1)\right]  \\ &+\frac{\partial_t\langle {\hat\rho}^2 \rangle_{in}}{\omega_0} \sin\left[\omega_0 (t-t_1)\right].
     \label{eq:solhom}
\end{aligned}
\end{align}
with $\omega_0$ from Eq.\eqref{eq:om0} and where we have used the following initial conditions:
\begin{equation}
\label{eq:con}
\begin{aligned}
    &\langle \hat\rho^2 \rangle_{in}=\langle \hat\rho^2 \rangle (t_1) = \langle \hat\rho^2 \rangle_0 +\langle \hat u_{\perp}^2 \rangle_0 t_1^2 ,\\ 
    &\partial_t\langle{\hat\rho}^2 \rangle_{in}= \partial_t\langle{\hat\rho}^2 \rangle (t_1) = 2\langle \hat u_{\perp}^2 \rangle_0 t_1,\\ 
    &\langle \hat u_{\perp}^2 \rangle (t_1) = \langle \hat u_{\perp}^2 \rangle_0,
    \end{aligned}
\end{equation}
and introduced the notation
\begin{equation}
    \langle \hat\rho^2 \rangle_{st} = \frac{2 \langle\hat{u}_{\perp}^2 \rangle_0 - \frac{2}{m} \omega_0 l}{\omega_0^2}.
    \label{eq:rstt}
\end{equation}
% Where as before $l=\langle \hat L_z \rangle$ is the eigenvalue of the operator of $z$-projection of the angular momentum.
In contrast to the previously considered solutions Refs.\cite{Karlovets2021,Silenko2021,FarEff}, Eq.\eqref{eq:solhom} contains an additional term, defined by the nonzero derivative $\partial_t\langle{\hat\rho}^2\rangle$ at the lens entrance.

We now recall the definition of the velocity operator Eq.\eqref{eq:vel} and express mean square of the velocity operator though the transverse energy of the Landau state Eq.\eqref{eq:LandauE} as
\begin{align}
    \langle\hat{u}_{\perp}^2 \rangle_0=\frac{\omega_0}{m}\left(2n'+|l| + l+1 \right),
\end{align}
and immediately arrive at
\begin{align}
   \langle \hat\rho^2 \rangle_{st} = \frac{\rho_H^2}{2}\left(2n'+|l|+1 \right),
   \label{eq:rsqLD}
\end{align}
in full agreement with the previous results from Ref.\cite{Karlovets2021}. 
We use the fact that $\langle\hat{u}_{\perp}^2 \rangle_0$ is continuous at the lens-vacuum interface, consequently, Eq.\eqref{eq:rstt} can be rewritten in terms of the transverse velocity of the free LG packet.
For the free LG packet, $\langle\hat{u}_{\perp}^2 \rangle_0$ reads \begin{align}
    \langle\hat{u}_{\perp}^2 \rangle_0=\frac{1}{m^2\sigma_r^2}\left(2n+|l|+1 \right).
    \label{eq:velfree}
\end{align}
with 
$$
\sigma^2_r\equiv \sigma^2_\perp(0)\equiv \langle \hat\rho^2\rangle(0).
$$
We note that the prefactor in Eq.\eqref{eq:rsqLD} can be also rewritten in terms of the electron Compton wavelength as $1/m^2/\sigma_r^2\equiv \lambda_c^2/\sigma_r^2$. 
Note that the coefficient 2 in front of $n$ was previously omitted in Refs.\cite{Karlovets_paraxial1,Karlovets2021}. We present the explicit calculations in the Appendix \ref{sec:2mult}.

Substituting Eq.\eqref{eq:velfree} into Eq.\eqref{eq:rstt} and expressing $\omega_0$ through $\rho_H$ with the help of Eq.\eqref{eq:rhom}, we get
\begin{align}
   \langle \hat\rho^2 \rangle_{st} = \frac{\rho_H^2}{2}\left[\frac{\rho_H^2}{4 \sigma_r^2} (2n+|l|+1) - l \right],
   \label{eq:rsqFR}
\end{align}
Equating left-hand sides of Eq.\eqref{eq:rsqLD} and Eq.\eqref{eq:rsqFR} we arrive at the condition on the ratio of the initial packet mean-square radius $\sigma_r$ and characteristic size of the Landau orbit  $\rho_H$
\begin{align}
    \frac{\rho_H^2}{\sigma_r^2}=4\frac{2n'+|l| + l+1}{2n+|l|+1},
    \label{eq:matchcond}
\end{align}
which is just the condition 
\begin{align}
 \label{eq:ucont}
\langle \hat u_{\perp}^2\rangle_{\text{free}} = \langle \hat u_{\perp}^2\rangle_{\text{lens}}.
\end{align}
  The conditions $\langle \hat\rho^2\rangle_{\text{free}} = \langle \hat\rho^2\rangle_{\text{lens}}, \partial_t\langle \hat{\rho}^2\rangle_{\text{free}} = \partial_t\langle \hat{\rho}^2\rangle_{\text{lens}}$ together with (\ref{eq:ucont}) provide continuation of the packet \textit{transverse emittance} \cite{Karlovets2021}
\begin{align}
 \label{eq:emcont}
\sqrt{\langle \hat\rho^2\rangle\langle \hat u_{\perp}^2\rangle - \langle \hat {\bm \rho}\cdot  \hat {\bm u}_{\perp}\rangle^2}_{\text{free}} = \cr = \sqrt{\langle \hat\rho^2\rangle\langle \hat u_{\perp}^2\rangle- \langle  \hat {\bm\rho}\cdot  \hat {\bm u}_{\perp}\rangle^2}_{\text{lens}}.
\end{align}
 
   Clearly, to fulfill the continuity of $\langle\hat{u}_{\perp}^2 \rangle_0$, the ratio $\rho_H^2/\sigma_r^2$ must be a rational number. This can be understood as a \textit{matching condition} between the \textit{free LG packet} and a  \textit{non-stationary Landau state} inside the lens. 
Another observation that follows from Eq.\eqref{eq:matchcond} is that if $n=n'$ and $|l|\gg 1$ the matching condition gets independent of any parameter except for the sign of $l$, namely
\begin{align}
    \frac{\rho_H^2}{\sigma_r^2}\approx 4[1+\mathrm{sgn} (l)].
    \label{eq:matchcond_simp}
\end{align}
Remarkably in this case we observe that if $l<0$ than $\sigma^2_r\gg \rho_H^2$ while for the case of $l>0$ matching condition becomes almost independent of $l$ and connects magnetic field and initial packet size as follows
\begin{align}
\label{eq:largeL}
    \sigma_r^2\approx \frac{\rho_H^2}{8}=\frac{1}{2e H_0}.
\end{align}
A quick estimate with the help of the Eq.\eqref{eq:largeL} gives $\sigma_r=573$nm for $H=10$G and $\sigma_r=18$nm for $H=10^4$G.
If the matching condition is met, the incoming LG packet can be captured by the lens. Inside the lens, the packet mean square radius $\langle \hat\rho^2 \rangle$ oscillates around the value defined by the Landau orbit given by Eq.\eqref{eq:rsqLD}, according to Eq.\eqref{eq:solhom}. 

\begin{figure}[t]
\center{\includegraphics[width=1.\linewidth]{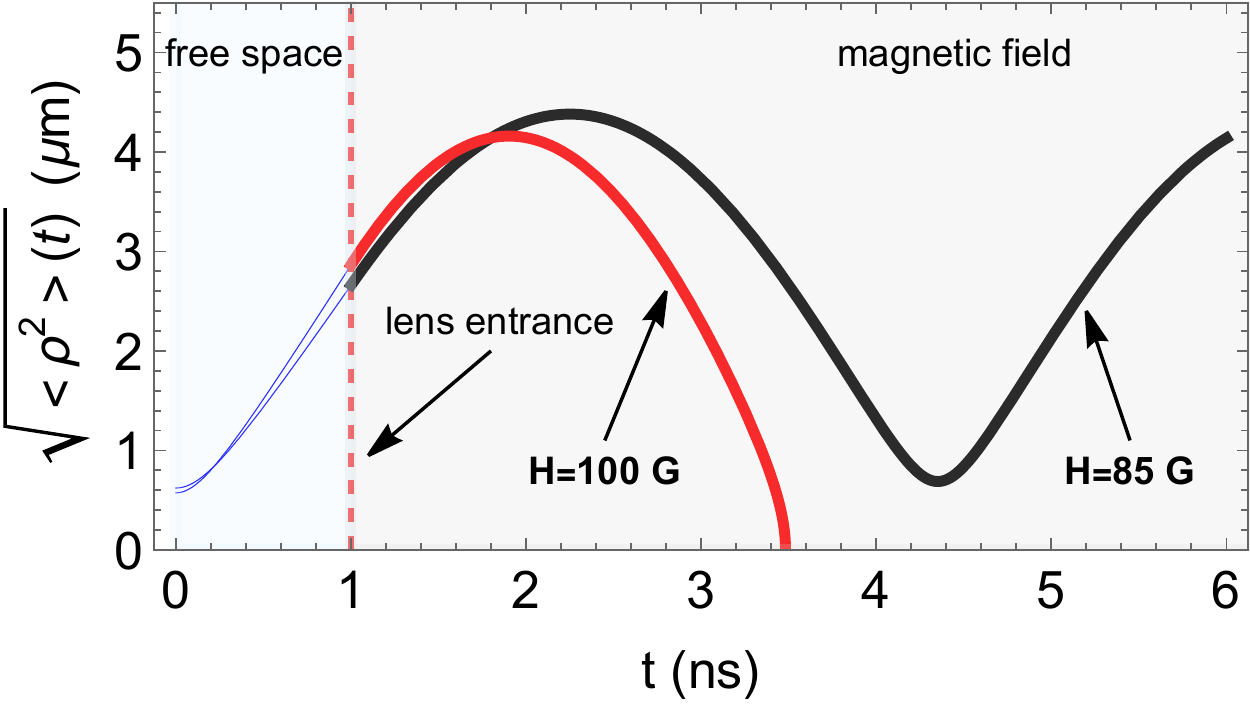}}
\caption{Evolution of the dispersion $\langle \hat\rho^2 \rangle(t)$ of the LG packet in the case of capture and transport (blue-red line) of the free LG packet and overfocusing (blue and black line). In free space (blue curve), $\langle \hat\rho^2 \rangle(t)$ is given by Eq.\eqref{eq:solFree}, inside the lens (black and red lines) by Eq. \eqref{eq:solhom}. The radial quantum number $n=n'=0$, the angular momentum $l=-4$ (following Eq.\eqref{eq:matchcond} we get $\rho_h^2/\sigma_r^2=0.8$); the magnetic field is $H=100$G with $\sigma_r=0.574\mu$m (redline) and $H=85$G with $\sigma_r=0.622\mu$m (black line). The initial square of the transverse velocity $\langle \hat u^2_\perp \rangle_0$ is defined with the help of the Eq.\eqref{eq:velfree}.}

\label{Fig:2}
\end{figure}

Interestingly, condition Eq.\eqref{eq:matchcond} is not sufficient to propagate the LG state successfully inside the lens. A brief inspection of the Eq.\eqref{eq:solhom} shows that $\langle \hat \rho^2 \rangle(t)$ is not necessarily positive for all moments in time even if the matching condition given by Eq.\eqref{eq:matchcond} is met. Consequently, additional \textit{transport condition} could be formulated by requesting $\langle \hat \rho^2 \rangle(t)>0$ and with the help of Eq.\eqref{eq:solhom} reads
\begin{align}
    \langle \hat \rho^2 \rangle_{st}-\sqrt{\left[ \langle \hat\rho^2 \rangle_{in} - \langle \hat\rho^2 \rangle_{st} \right]^2+\left[\frac{\partial_t\langle{\hat\rho}^2 \rangle_{in}}{\omega_0} \right]^2}>0.
    \label{eq:trcond0}
\end{align}
Solving Eq.\eqref{eq:trcond0} with respect to $ \langle \hat \rho^2 \rangle_{st}$ we get 
\begin{align}
    \langle \hat \rho^2 \rangle_{st}>\frac{\langle \hat \rho^2 \rangle_{in}}{2}+\frac{\left[\partial_t\langle{\hat\rho}^2 \rangle_{in}\right]^2}{2\omega_0^2\langle \hat \rho^2 \rangle_{in}}.
\label{eq:trcond}
\end{align}
We note that for the case of $\partial_t\langle{\hat\rho}^2 \rangle_{in}=0$ this condition reduces to the one mentioned in Ref.\cite{Karlovets2021}.
The \textit{transport condition} (for a given OAM of the free LG packet) sets limits on the magnetic field strength, radial quantum number and a drift time of the free LG packet before it enters the lens. Examples of the possible evolution scenarios are plotted in Fig.\ref{Fig:2}. One can see from the figure that if the transport condition is not met, then the mean square radius $\langle \hat \rho^2 \rangle$ inside the lens rapidly becomes negative and the dispersion becomes complex. This in turn indicates the breakdown of the developed formalism and, most likely, destruction of the pure LG state inside the lens because focusing to the spot $\langle \hat \rho^2 \rangle < \lambda_c^2$ implies violation of the one-particle dynamics with a stable vacuum \cite{BLP}.  % We may even assume that in this case the packet is overfocused and most probably becomes delocalized.

Longitudinal dynamics for the average value of the longitudinal momentum and coordinate is trivial and easily recovered from the corresponding Heisenberg's equation Eq.\eqref{eq:Heis} for the corresponding observable. We omit simple derivations and provide the final result for $t>t_1$  
\begin{align}
    &\langle \hat p_z \rangle = p_0+e |E_0| (t-t_1), \nonumber\\
    &\langle \hat z \rangle= \frac{p_0}{m}(t-t_1)+\frac{e |E_0|}{m} \frac{(t-t_1)^2}{2}.
\label{eq:longM}
\end{align}
that we are going to use in the further analysis. Above and further for the sake of convenience we shifted the $z$ axis such that $\langle \hat{z} \rangle(t_1)=0.$ 

As expected, average of the $z$ - projection of the momentum operator $\langle \hat p_z \rangle$ as well as the average of the $z$ - coordinate operator $\langle \hat z \rangle$ is described by a uniformly accelerated motion.

%================================================================================================
\subsection{Inhomogeneous lens \label{sec:IHL}}
%================================================================================================

%\textcolor{red}{Vse, chto bylo do etogo razdela bolee-menee izvestno. Etot razdel soderzhit novelty. Poetomu predlagayu napisat ego v bolee accessible forme, dobavit slova i poyasneniya.}

Now we consider full Hamiltonian given by Eq.\eqref{eq:lensHam}. Accounting for Eq.\eqref{eq:pots} and evaluating commutators on the left hand side of the system Eq.\eqref{eq:mainsys}, we get
\begin{equation}
\label{eq:inhsys}
\left\{\begin{aligned}
    &\frac{d^2 \hat \rho^2}{dt^2} = \frac{2\hat{p}_{\perp}^2}{m^2} -\left[ \frac{\omega^2_c(\hat z)}{2}-\frac{e}{m} E_1 \right] \hat\rho^2 , \\
    & \frac{d \hat{u}_{\perp}^2 }{d t} =\frac{\kappa_M \omega_0}{m^2 L}\left( \hat L_z + \frac{m\omega_0}{2} \hat\rho^2\right)\hat{p}_z + \\ & + (\hat{p}_z \hat z + \hat z \hat{p}_z)
    \frac{\kappa_M^2 \omega_0^2}{4 m L^2} \hat \rho^2,
\end{aligned}\right.
\end{equation}
where $\omega_0$ and $\kappa_M$  are given by Eq.\eqref{eq:om0} and Eq.\eqref{eq:aM}, respectively.

Expressing $\hat p_\perp$ via the $\hat u_\perp$ with the help of Eq.\eqref{eq:vel} and the following definition of the transverse and longitudinal Hamiltonians 
\begin{align}
\label{eq:prLHfl}
    \hat{H}_\perp=& \frac{\hat{p}_\perp^2}{2m}+ \frac{\omega_c(\hat z)\hat L_z}{2}+ \left[\frac{m\omega_c(\hat z)^2}{8}-\frac{eE_1}{4}\right] \hat \rho^2,
\end{align}
and 
\begin{align}
\label{eq:lnLHfl}
    \hat{H}_\parallel=\frac{\hat p_z^2}{2m}-e |E_0| \hat z+\frac{eE_1}{2}\hat z^2,
\end{align}
we average over the system state and finally get
\begin{equation}
\label{eq:inhsysav}
\left\{\begin{aligned}
    &\frac{d^2 \langle \hat\rho^2 \rangle}{d t^2} = 2 \langle \hat{u}_{\perp}^2 \rangle \\&- \frac{2}{m} \langle \omega_c(\hat z) \rangle l - \left\langle \left( \omega_c^2(\hat z) - \frac{2 e}{m} E_1 \right) \hat \rho^2 \right\rangle, \\
    & \frac{d \langle \hat{u}_{\perp}^2 \rangle}{d t} = \frac{\kappa_M \omega_0}{m^2 L}\left(l \langle \hat{p}_z \rangle + \frac{m\omega_0}{2} \langle \hat\rho^2 \hat{p}_z\rangle\right)+\\ & + \left(\langle \hat{p}_z \hat z\hat \rho^2 \rangle+\langle \hat z \hat{p}_z \hat\rho^2\rangle \right)
    \frac{\kappa_M^2 \omega_0^2}{4 m L^2}.
 \end{aligned}\right.
\end{equation}
Note that for inhomogeneous fields the longitudinal and transverse Hamiltonians, given by Eq.\eqref{eq:prLHfl} and Eq.\eqref{eq:lnLHfl}, do not commute anymore $\left[\hat H_\perp,\hat H_\parallel \right]\neq 0$ as separation of the variables is no longer possible. This results in the correlation between the longitudinal and transverse operators and an average of the following type cannot be factorised 
\begin{equation}
    \langle \Psi | \hat{A}(\rho)\hat{B}(z) | \Psi \rangle \ne \langle \Psi | \hat{A}(\rho) | \Psi \rangle \langle \Psi | \hat{B}(z) | \Psi \rangle.
\end{equation}
As a consequence Eqs.\eqref{eq:inhsysav} is not a closed system of equations, moreover, one can check that the system won't close even if the order of the time derivative is raised and higher order commutators on the right hand side are evaluated.

To proceed further, we consider \textit{inhomogeneity as a small perturbation} to the homogeneous problem. We introduce the following small parameters of the problem:
\begin{align}
    &\kappa_M=\frac{H_1 L}{H_0} \ll 1, \\ \nonumber 
    &\kappa_E=\frac{E_1 L}{E_0} \ll 1,
\end{align}
where $L$ is a characteristic length (it can be the length of the lens). 

We assume that $\kappa_M \sim \kappa_E$ and decompose the wave function in series with respect to the small parameter $\kappa=\kappa_E=\kappa_M$
\begin{align}
    \ket{\Psi}=\ket{\Psi}^{(0)}+\ket{\Psi}^{(1)}+\mathcal{O}(\kappa^2).
\end{align}
Here $\ket{\Psi}^{(0)}$ is the non-stationary wave function of the homogeneous lens, which corresponds to the RMS given by Eq.\eqref{eq:solhom} satisfying the system Eq.\eqref{eq:mainsysFL}. The correction $\ket{\Psi}^{(1)}$ is  due to gradients of the fields and naturally $\left| \left| \, \left| \Psi^{(1)} \right\rangle \right| \right|\sim \kappa \ll \left|  \left| \, \left|\Psi^{(0)} \right\rangle \right| \right|=1$. This leads to the following series for the averages:
\begin{align}
    &\langle \hat \rho^2 \rangle=\langle \hat\rho^2 \rangle^{(0)} + \langle \hat\rho^2 \rangle^{(1)}+\mathcal{O}(\kappa^2),\\
    & \langle \hat{u}_{\perp}^2 \rangle = \langle \hat{u}_{\perp}^2 \rangle^{(0)}  + \langle \hat{u}_{\perp}^2 \rangle^{(1)}+\mathcal{O}(\kappa^2),
\end{align}
where

\begin{align}
    &\langle \hat \rho^2 \rangle^{(0)}=\left\langle \Psi^{(0)} \right| \hat \rho^2 \left| \Psi^{(0)} \right\rangle, \nonumber \\ 
    &\langle \hat{u}_{\perp}^2 \rangle^{(0)}=\left\langle \Psi^{(0)} \right| \hat{u}_{\perp}^2 \left( E_1, H_1 = 0 \right) \left| \Psi^{(0)} \right\rangle, \nonumber\\
    &\langle \hat \rho^2 \rangle^{(1)}=\left\langle \Psi^{(1)} \right| \hat \rho^2 \left| \Psi^{(0)} \right\rangle+\left\langle \Psi^{(0)} \right|\hat \rho^2 \left| \Psi^{(1)} \right\rangle, \\
    &\langle \hat{u}_{\perp}^2 \rangle^{(1)}=\left\langle \Psi^{(1)} \right|\hat \rho^2 \left| \Psi^{(0)} \right\rangle+\left\langle \Psi^{(0)} \right| \hat{u}_{\perp}^2 \left| \Psi^{(1)} \right\rangle + \nonumber \\
    & + \left\langle \Psi^{(0)} \right| \hat{u}_{\perp}^2 - \hat{u}_{\perp}^2 \left( E_1, H_1 = 0 \right) \left| \Psi^{(0)} \right\rangle. \nonumber
\end{align}
In the first order in $\kappa$, the system Eq.\eqref{eq:inhsysav} reads
\begin{equation}
\label{eq:syscorF}
\left\{\begin{aligned}
    &\frac{d^2 \langle \hat \rho^2 \rangle^{(1)}}{d t^2} + \omega_0^2 \langle \hat \rho^2 \rangle^{(1)}= 2 \left\langle \hat{u}_{\perp}^2 \right\rangle^{(1)} - \kappa \frac{2 \omega_0 l}{m L} \langle \hat z \rangle^{(0)}\\ & - \kappa \frac{2 \omega_0^2}{L} \langle \hat z \rangle^{(0)} \langle \hat \rho^2 \rangle^{(0)}+ \kappa \frac{2 e E_0}{m L} \langle \hat \rho^2 \rangle^{(0)}, \\
    & \frac{d \langle \hat{u}_{\perp}^2 \rangle^{(1)}}{d t} = \frac{\kappa \omega_0}{m^2 L}\left( l + \frac{m\omega_0}{2} \langle \hat\rho^2 \rangle^{(0)}\right)\langle\hat{p}_z\rangle^{(0)}.
\end{aligned}\right.
\end{equation}
and it should be complemented with the initial conditions at the lens entrance, 
\begin{align}
    \langle \hat{u}_{\perp}^2 \rangle^{(1)}(t_1)=0, \nonumber \\
    \langle \hat \rho^2 \rangle^{(1)}(t_1)=0, \\
    \frac{d\langle \hat \rho^2 \rangle^{(1)}}{dt}(t_1)=0. \nonumber
\end{align}
Explicit formulas for all approximations and assumptions could be found in the Appendix \ref{sec:app3}.
\begin{widetext}

After some algebra, the final expression for the correction $\langle \hat \rho^2 \rangle^{(1)}$ that satisfies the system Eq.\eqref{eq:syscorF} reads
\begin{align}
\label{eq:corr}
\langle \hat \rho^2 \rangle^{(1)}=& -\frac{\kappa \sin \left[\omega_0 \Delta t \right]}{2 L m \omega_0} \left \{e E_0
   \Delta t \left[\partial_t\langle{\hat\rho}^2 \rangle_{in} \Delta t-4 \left(\langle \hat \rho^2 \rangle_{in}- \langle \hat \rho^2 \rangle_{st} \right) \right]+2 p_0 \left[\partial_t\langle{\hat\rho}^2 \rangle_{in} \Delta t-\langle \hat \rho^2 \rangle_{in}\right]\right\}  \\
   &  -\frac{\kappa \sin \left[\omega_0 \Delta t \right]}{6 L m \omega_0^3}\left\{\omega_0^4 \Delta t^2 \left[\langle \hat \rho^2 \rangle_{in}-\langle\hat \rho^2 \rangle_{st}\right]  \left[e E_0
    \Delta t+3 p_0\right]-12 e E_0 \partial_t\langle{\hat\rho}^2 \rangle_{in}\right\} \nonumber\\
    &+\frac{\kappa \cos
   \left[ \omega_0 \Delta t \right]}{L m \omega_0^2} \left\{e E_0
    \left[\langle \hat \rho^2 \rangle_{in}-4\langle \hat \rho^2 \rangle_{st}-2\partial_t\langle{\hat \rho}^2 \rangle_{in} \Delta t\right]-p_0 \partial_t\langle{\hat \rho}^2 \rangle_{in}\right\} \nonumber \\ 
    &+\frac{\kappa \cos
   \left[ \omega_0 \Delta t \right] \Delta t}{6 L m}\left\{e E_0 \Delta t \left[\partial_t\langle{\hat \rho}^2 \rangle_{in} \Delta t -3 \left( \langle \hat \rho^2 \rangle_{in}-\langle \hat \rho^2 \rangle_{st} \right) \right]+3p_0 \left[\partial_t\langle{\hat \rho}^2 \rangle_{in} \Delta t -2 \left(\langle \hat \rho^2 \rangle_{in}-\langle \hat \rho^2 \rangle_{st} \right) \right] \right\}  \nonumber\\
   &-\frac{\kappa}{2 L m \omega_0^3}  \left\{2 \omega_0 \left[eE_0
   \left(\langle \hat \rho^2 \rangle_{in}-4 \langle \hat \rho^2 \rangle_{st} \right)-p_0 \partial_t\langle{\hat \rho}^2 \rangle_{in}
   \right]+\langle \hat \rho^2 \rangle_{st} \omega_0^3
   \Delta t \left[eE_0 \Delta t+2p_0 \right])\right\} \nonumber.
\end{align}

\end{widetext}

\begin{figure}[b]
\center{\includegraphics[width=1.\linewidth]{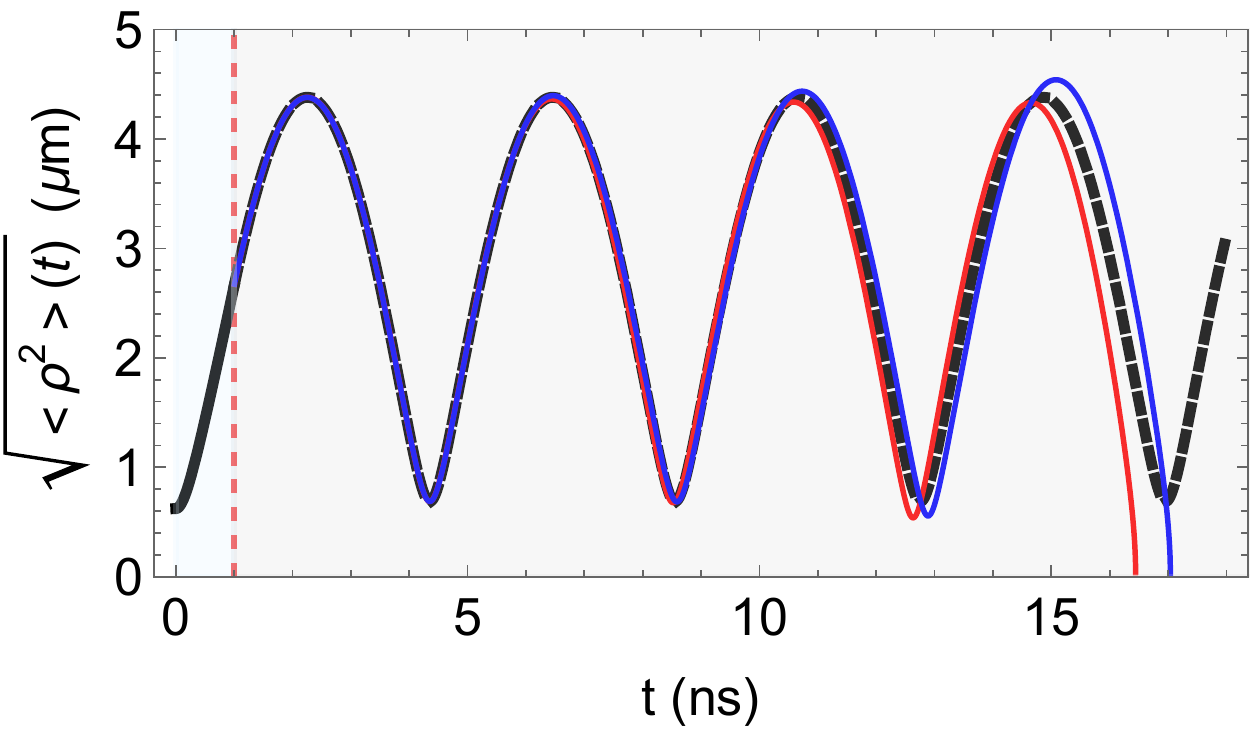}}
\caption{Evolution of the dispersion $\langle \hat\rho^2 \rangle(t)$ of the LG packet in the case $\kappa=0$ - black dashed line line, $\kappa=0.081$ - red line, $\kappa=-0.081$ blue line.}

\label{Fig:2c}
\end{figure}

Here $\Delta t \equiv t-t_1$, $L$ is the full length of the lens, $\langle \hat \rho^2 \rangle_{in}$ and $\partial_t\langle{\hat \rho}^2 \rangle_{st}$ are given by Eq.\eqref{eq:con}; $\langle \hat \rho^2 \rangle_{st}$ is defined in Eq.\eqref{eq:rsqLD}. 

In Fig.\ref{Fig:2c} we plot $\sqrt{\langle \hat{\rho}^2 \rangle}$ for three different cases: $\kappa=0$, $\kappa=0.081$ and $\kappa=-0.081$. As before, the parameters are as follows: initial packet size $\sigma_r=0.622 \mu$m, radial quantum number $n=n'=0$, angular momentum $l=-4$, magnetic field $H=85$G and initial longitudinal momentum $p_0=0.43$eV.
We observe that inhomogeneity alters the motion just slightly at the beginning and is observable as the frequency shift as well as the slight growth of the oscillation amplitude. Interestingly in both cases after some propagation distance (four periods in the considered case) oscillations break down and over-focusing occurs. We conclude that inhomogenuity affects the life time of the twisted state and, non the less $15$ ns is a large time scale in a real life experiment, should be accounted for as it may constrain acceptable tolerances of the experimental setup.

%================================================================================================
\section{Discussion}
%================================================================================================
The developed formalism allows one to gain an insight into the parameters needed for acceleration and transport of charged particles with the phase vortices. For instance, to get an idea of what the physical dimensions of the experimental setup could be we first estimate the drift length as follows
\begin{align}
    L_{\text{free}}=\frac{\langle \hat p_z \rangle_0}{m}t_1.
\end{align}
We consider parameters of a toy model introduced earlier in Fig.\ref{Fig:2}. If we take thermionic emission with the source temperature of $5000$K, then an equivalent momentum is roughly $\langle \hat p_z \rangle_0\sim 0.43$eV. Consequently, in $t_1=1$ns the drift distance becomes $L_{\text{free}}=0.253\mu$m. 
On the other hand, roughly one oscillation inside the lens occurs within $\Delta t\sim 6$ns. The accelerating field that is routinely achieved inside a linear accelerator cavity is $E_z\sim 25$ MV/m \cite{TESLA}.  According to the Eq.\eqref{eq:longM}, $3$ ns inside the lens with the initial conditions outlined above transforms into $L_{\text{lens}}\sim 38$ m. The latter estimate must be treated with caution as at this point the particle becomes ultra-relativistic and to get a proper estimate relativistic effects must be included in the considerations, which will be studied elsewhere. Nonetheless, we observe that drift distance is very small compared to the acceleration distance and this in turn may result in practical limitations. To increase the drift distance and meet the transport condition one should relax the magnetic field, this, in turn, will result in just a few oscillations inside the lens indicating that the process is potentially always transitional and will never reach an equilibrium stationary Landau state given by Eq.\eqref{eq:Landau}.  

We point out that inside the lens $\partial_t^3\langle \hat \rho \rangle\neq 0$ in contrast to free space where as can be seen from Eq.\eqref{eq:solFree} $\partial_t^3\langle \hat \rho \rangle\neq 0$. Condition $\partial_t^3\langle \hat \rho \rangle\neq 0$ result in a chance in of the electric quadrupole moment and consequent radiation and dissipation of the particle energy. This could be interpreted as an analog of a synchrotron radiation. The full dynamics could be understood as follows. An expanding LG wave packet enters the lens and exhibits oscillation of the $\langle \hat \rho^2 \rangle$. Oscillation of the $\langle \hat \rho^2 \rangle$ results in the oscillation of the electric quadrupole moment of the packet and consequently the packet radiates up to the point when the state reaches an equilibrium that is described by the stationary Landau state Eq.\eqref{eq:Landau}. The radiation rate and consequently relaxation constant of this process could be estimated via the intensity of the quadrupole moment radiation and is proportional to $\partial_t^3 \langle \hat \rho^2 \rangle \sim \omega_0^3\sim H_0^3$. 

\begin{figure}[t]
\center{\includegraphics[width=1.\linewidth]{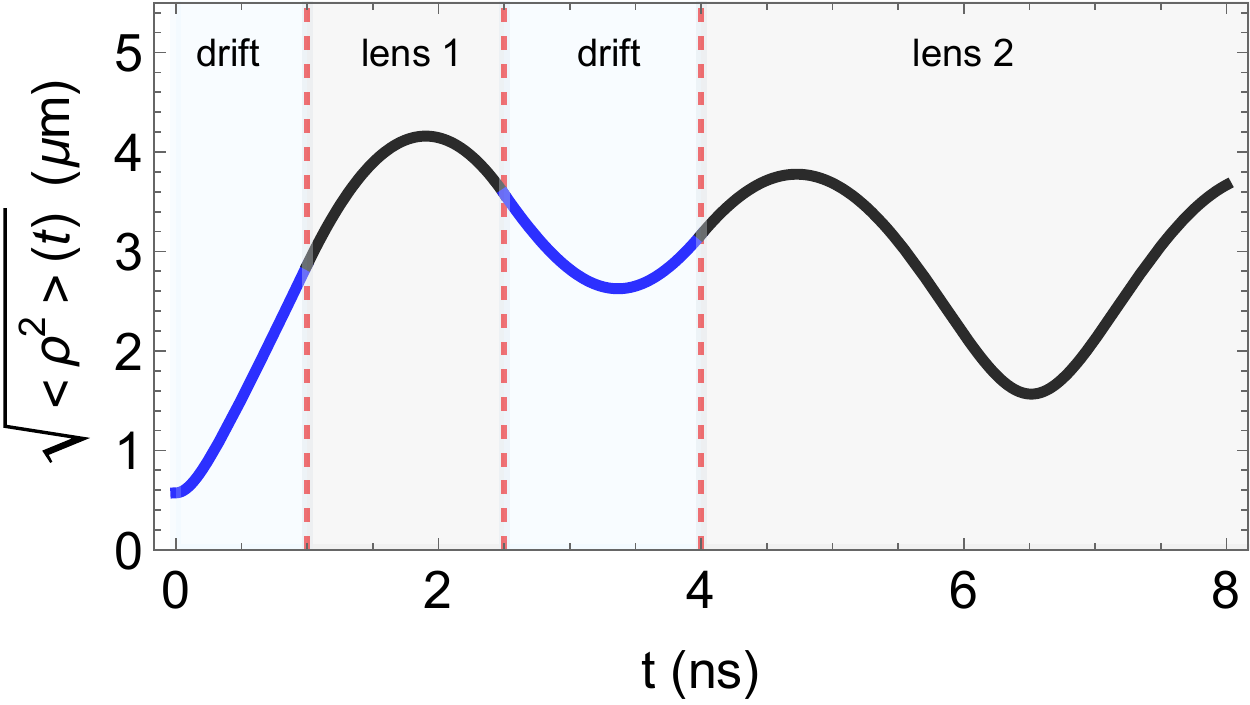}}
\caption{Evolution of the dispersion $\langle \hat\rho^2 \rangle(t)$ of the LG packet for two lenses separated by a drift. Due to the packet's excessive size and $H=100$G transport condition Eq.\eqref{eq:trcond} for the first lens is broken and the packet is over-focused. Before the over-focusing develops the first lens ends at $t=2.5$ns and the packet evolves according to the Eq.\eqref{eq:solFree}; at $t=4$ns the transport condition Eq.\eqref{eq:trcond} for the second lens (which is identical to the first one) is fulfilled. The second lens captures the packet at $t=4$ns and transports the packet further.}
\label{Fig:3}
\end{figure}

The formalism presented in Sec.\ref{sec:HL} is extendable to \textit{any number of drifts and magnetic lenses} that are combined in a lattice. This in turn allows one to operate each individual lens in the over-focusing regime (if needed) and still get a stable propagation of an LG packet through the combined channel. Indeed, let's again consider the toy model as described in Fig.\ref{Fig:2}. We recall the parameters: initial packet size was $\sigma_r=0.574 \mu$m, radial quantum number $n=0$, angular momentum $l=-4$ and magnetic field $H=100$G (over-focusing regime). As before initial square of the transverse velocity $\langle \hat u^2_\perp \rangle_0$ is defined with the help of the Eq.\eqref{eq:velfree}. Terminating the first lens at $t=2.5$ ns and setting the second lens with the same $H=100$G at $t=4$ns we are able to fulfill the transport condition for the combined system and the packet can successfully propagate inside the second lens (see Fig.\ref{Fig:3} for details). It is important to keep in mind that the matching condition must be fulfilled as well. In the considered case in the second drift, the packet was expanded such that dispersion at the focal point $t=3.367$ ns is increased and about 21 times larger than the initial value of $0.574 \mu$m. To fulfill the matching condition $\rho_h^2/\sigma_r^2/21 = 0.8/21=2/210$ the packet must have a radial quantum number $n=50$ (according to the Eq.\eqref{eq:matchcond} with $n'=0$) in the second drift. This implies that after exiting the second lens the packet must undergo the following transition $(n,l,p_z):(0,-4,p_z)\to(50,-4,p_z)$. While this transition is possible it is quite likely to have a low probability. Detailed analysis of this question is the focus of a separate study.         

\begin{figure}[t]
\center{\includegraphics[width=1.\linewidth]{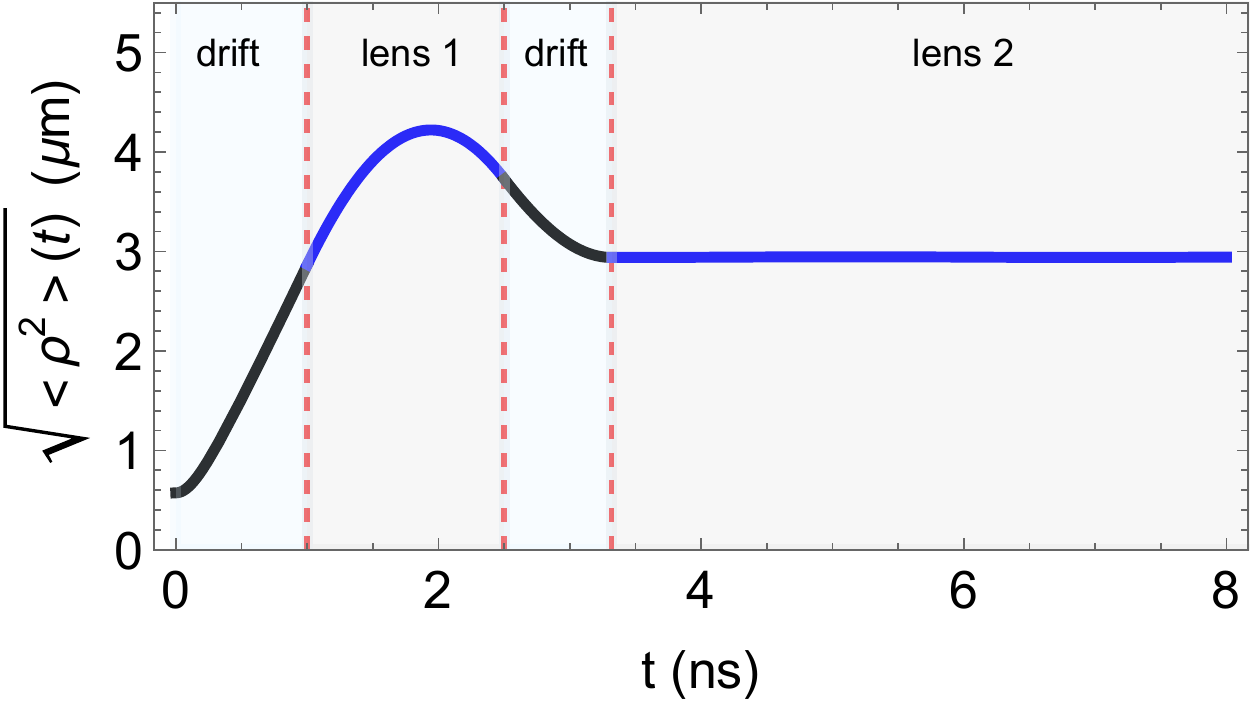}}
\caption{Evolution of the dispersion $\langle \hat\rho^2 \rangle(t)$ of the LG packet for two lenses separated by a drift. The second lens is placed at the focal point such that after the second drift a direct capture of the free wave packet into the stationary Landau state occurs.}
\label{Fig:5}
\end{figure}

Interestingly, with the set of the magnetic lens, any free packet could be directly captured into a \textit{stationary Landau state}. As discussed in Sec.\ref{sec:2} this process is impossible with just one lens. If the second lens is placed at the focal point, where the time derivative $\partial_t \langle \hat \rho^2 \rangle=\partial_t\langle{\hat\rho}^2 \rangle_{in} =0$ and tune the magnetic field of the second solenoid such that the RMS radius of the stationary Landau orbit matches the size of the incoming packet $\langle \hat\rho^2 \rangle_{st}=\langle \hat\rho^2 \rangle_{in}$, then according to the Eq.\eqref{eq:solhom} no oscillations occur and the state is a pure stationary Landau state given by Eq.\eqref{eq:Landau}. To illustrate this case we pick the same parameters as previously except for the magnetic field of solenoids that is now $H=97.8$G and the second lens is placed right at the point $t=3.32$ ns where $\partial_t \langle \hat \rho^2 \rangle=0$. The result is displayed in Fig.\ref{Fig:5}. We note that this case is quite special and such matching could be very challenging in a realistic experiment. 

Although the present study unveils rich dynamics and demonstrates several concepts, it is still worthy to investigate the problem from a more rigorous perspective of quantum scattering to confirm the following: breakdown of the matching condition Eq.\eqref{eq:matchcond} may lead to reflection of the free packet, whereas breakdown of the transport condition Eq.\eqref{eq:trcond} may lead to the rapid delocalization of the LG state inside the lens. We would like to stress that in the present study we have considered only the direct transition of the pure state of the free LG packet to the pure nonstationary state in the magnetic field. It is quite possible that the capture mechanism and consequent dynamics get more complex once the state inside the lens becomes a superposition of the pure states. 

\section{Conclusion}

We have analyzed the acceleration of a nonrelativistic charged particle with a phase vortex in the inhomogeneous magnetic lens. In free space and inside the homogeneous magnetic field, the electron can be described via the standard LG packet, which is an exact solution to the nonstationary Schr\"{o}dinger equation. The time evolution of such a state is described by the time dependence of the mean square radius and the phase. Whereas the OAM is conserved by neglecting the radiation, the spatial structure of the LG packet is preserved during the transport only if the matching condition, given by Eq.\eqref{eq:matchcond}, is fulfilled. Moreover, if the transport condition given by Eq.\eqref{eq:trcond} is also fulfilled, the free LG packet can be directly transformed into the nonstationary Landau state and propagated inside the magnetic lens with the conserved OAM. We hypothesize a possible mechanism of further evolution of the \textit{nonstationray Landau state} into the \textit{stationary Landau state} by means of the radiation friction caused by oscillations of the electric quadrupole moment of the packet inside the lens.    

A toy model of a possible experimental setup is considered, and preliminary calculations demonstrate the reasonable feasibility of the potential experiment.
Finally, we have analyzed the effect of inhomogeneity of the electromagnetic fields and derived analytical expressions Eq.\eqref{eq:corr} for the corrections to the evolution of the packet mean square radius. 
Further developments in this direction imply relativistic considerations where the above quantum scattering at the boundary and destruction of the Landau state will likely play a role.

\begin{acknowledgments}
The studies in Sec.II are supported by the Russian Science Foundation (Project No. 17-72-20013), in Sec.III by the Russian Science Foundation (Project No. 22-22-20062), and in Sec. IV by the government of the Russian Federation through the ITMO Fellowship and Professorship Program.
\end{acknowledgments}

%++++++++++++++++++++++++++++
\appendix
%++++++++++++++++++++++++++++

\section{Maxwell equations for the electric and magnetic fields. Check by the substitution.}
\label{app:1}
%================================================================================================
We consider divergence operator in cylindrical coordinates 
\begin{align}
\label{eq:div}
 \mathrm{div}~\bm{\mathrm{f}}\equiv \frac{1}{\rho}\frac{\partial \rho f_\rho}{\partial \rho} +\frac{1}{\rho}\frac{\partial f_\phi}{\partial \phi}+\frac{\partial f_z}{\partial z}
\end{align}
and curl operator in cylindrical coordinates
\begin{align}
 \label{eq:rot}
    \mathrm{rot}~\mathbf{f} &\equiv \mathbf{e_\rho} \left(\frac{\partial_\phi f_z}{\rho}-\partial_z f_\phi \right)+\nonumber \\&+\mathbf{e_\phi} \left(\partial_z f_\rho -\partial_\rho f_z \right)+\mathbf{e_z} \frac{1}{\rho}\left(\partial_\rho \left[ \rho f_\phi \right]-\partial_\phi f_\rho \right).  
\end{align}
We substitute Eq.\eqref{eq:elf} into Eq.\eqref{eq:div} and get
\begin{align}
    \mathrm{div}\bm{\mathrm{E}}=\frac{1}{\rho}\frac{\partial \rho \left[-\frac{E_1}{2} \rho\right]}{\partial \rho} +\frac{\partial \left[E_{ac}+E_1 z\right]}{\partial z}=0.
\end{align}
Same for the magnetic field, with Eq.\eqref{eq:magf} and Eq.\eqref{eq:div} we get
\begin{align}
    \mathrm{div}\bm{\mathrm{H}}=\frac{1}{\rho}\frac{\partial \rho \left[-\frac{H_1}{2} \rho\right]}{\partial \rho} +\frac{\partial \left[H_0+H_1 z\right]}{\partial z}=0.
\end{align}
As far as $E_\phi=0$ and $H_\phi=0$ as well as all $\partial_\phi$ derivatives are equal to zero we immediately arrive at 
\begin{align}
\label{eq:MXz}
    \mathrm{div}\bm{\mathrm{E}}=\mathcal{O}\left( \rho, z \right), \; \mathrm{rot}\bm{\mathrm{E}}=\mathcal{O}\left( \rho, z \right),\nonumber \\
    \mathrm{div}\bm{\mathrm{H}}=\mathcal{O}\left( \rho, z \right), \;\mathrm{rot}\bm{\mathrm{H}}=\mathcal{O}\left( \rho, z \right).
\end{align}

\section{Evaluation of commutators \label{sec:app2}}
\label{app:2}
%================================================================================================
We provide evaluation of main commutators that we utilized in the manuscript.

We make use of the following formulas.
\begin{align}
    &\left[F(\partial_{\rho}), \rho\right]= \frac{\partial F(x)}{\partial x} \bigg|_{x = \partial_{\rho}}, \\
    &\left[G(\rho), \partial_{\rho} \right]= - \frac{\partial G(x)}{\partial x} \bigg|_{x = \partial_{\rho}}, \\
    &\left[F(\partial_{\rho}),\rho^2\right]= 2 \rho \frac{\partial F(x)}{\partial x} \bigg|_{x = \partial_{\rho}} + \frac{\partial^2 F(x)}{\partial x^2} \bigg|_{x = \partial_{\rho}}, \\
     &\left[G(\rho), \partial^2_{\rho} \right]= - 2 \frac{\partial G(x)}{\partial x} \bigg|_{x = \partial_{\rho}} \partial_{\rho} - \frac{\partial^2 G(x)}{\partial x^2} \bigg|_{x = \partial_{\rho}}.
\end{align}

One can calculate the commutator of radius squared with Hamiltonian
\begin{align}
&\left[\hat{H},\rho^2\right]=\frac{1}{2m}\left[\hat{\bm{p}}^2,\rho^2\right]= \nonumber\\
&\frac{1}{2m}\left[ - \partial^2_{\rho} - \frac{1}{\rho} \partial_{\rho},\rho^2\right ] = - \frac{1}{2m}\left[ \partial^2_{\rho},\rho^2\right]-\frac{1}{m} = \\
& - \frac{1}{2m} \partial_{\rho} \left[ \partial_{\rho} ,\rho^2\right] - \frac{1}{2m}\left[ \partial_{\rho}, \rho^2\right] \partial_{\rho} - \frac{1}{m} = \nonumber \\
& -\frac{2 }{m}\rho \partial_{\rho} -\frac{2}{m}\nonumber.
\end{align}
Constant terms play no role in the following commutations so we neglect them.
\begin{align}
    &\left[\hat{H},\left[\hat{H},\rho^2\right]\right]=-\frac{2}{m}\left[\hat{H},\rho \partial_{\rho} \right]= \nonumber\\
    &-\frac{1}{m^2}\left[ - \partial^2_{\rho} -\frac{1}{\rho} \partial_{\rho} ,\rho \partial_{\rho} \right] + \frac{1}{m^2}\left[ \frac{1}{\rho^2} \partial^2_{\phi},\rho \partial_{\rho} \right]\\
    &- \frac{\omega^2_c(z)}{4}\left[\rho^2,\rho \partial_{\rho} \right]-\frac{eE_1}{m}[\rho^2,\rho \partial_{\rho} ]= \nonumber \\
    &\frac{2}{m^2} \partial^2_{\rho} +\frac{2}{m^2\rho} \partial_{\rho} + \frac{2}{m^2 \rho^2} \partial^2_{\phi} +\frac{\omega^2_c(z)}{2}\rho^2+\frac{2eE_1}{m}\rho^2= \nonumber \\
    &-2\hat{u}^2_{\perp} + \frac{2}{m} \omega_c(z) \hat{L}_z + \left( \omega_c^2(z) - \frac{2e}{m} E_1 \right) \rho^2\nonumber
\end{align}
Commutator of the square of the velocity operator with the lens Hamiltonian is evaluated as  
\begin{align}
    \left[\hat{H},\hat{u}^2_{\perp}\right]=\frac{2}{m}\left[\hat{H}_{\perp}+\hat{H}_\parallel,\hat{H}_{\perp}\right].
\end{align} 
Here $\hat{H}_\perp$ and $\hat{H}_\parallel$ are given by Eq.\eqref{eq:prLHfl} and Eq.\eqref{eq:lnLHfl} respectively.
Simplification gives
\begin{align}
    &\left[\hat{H},\hat{u}^2_{\perp}\right]=\frac{2}{m}\left[\hat{H}_\parallel,\hat{H}_{\perp}\right]= \nonumber\\
    &\frac{2}{m}\left[\frac{\hat{p}_z^2}{2m},\frac{\omega_c(\hat{z}) \hat{L}_z}{2}+\frac{m \omega_c(\hat z)^2}{8}\hat \rho^2\right]= \\ \nonumber
    &\frac{\kappa_M \omega_0}{2 m^2 L}\left( \hat L_z + \frac{m\omega_0}{2} \hat\rho^2\right) \left[\hat{p}_z^2,\hat z\right]+ \\
    &\frac{\kappa_M^2 \omega_0^2}{8 m L^2} \hat \rho^2\left[\hat{p}_z^2,\hat{z}^2\right].
\end{align} 
and finally we get 
\begin{align}
    &\left[\hat{H},\hat{u}^2_{\perp}\right]=
    -i\frac{\kappa_M \omega_0}{m^2 L}\left( \hat L_z + \frac{m\omega_0}{2} \hat\rho^2\right)\hat{p}_z - \\ \nonumber& i (\hat{p}_z \hat z  + \hat z \hat{p}_z)
    \frac{\kappa_M^2 \omega_0^2}{4 m L^2} \hat \rho^2.
\end{align}

\section{Approximations \label{sec:app3}}
%================================================================================================
In this appendix we explicitly list all approximations that were used to derive system \eqref{eq:syscorF} 
\begin{align}
    & \langle \omega_c^2(z) \rho^2 \rangle \approx \omega_0^2 \langle \rho^2 \rangle^{(0)} + \nonumber \\ &\left( \omega_0^2 \langle \rho^2 \rangle^{(1)} + 2 \omega_0 \omega_1 \langle z \rangle^{(0)} \langle \rho^2 \rangle^{(0)} \right), \\
    & \kappa \left\langle \rho^2 \hat{p}_z^2 \right\rangle \approx \kappa \left\langle \rho^2 \right\rangle^{(0)} \left\langle \hat{p}_z^2 \right\rangle^{(0)}, \\
    & \nonumber \frac{\kappa_M \omega^2_0}{2 m L} \langle \hat\rho^2 \hat{p}_z\rangle + \left(\langle \hat{p}_z \hat z\hat \rho^2 \rangle+\langle \hat z \hat{p}_z \hat\rho^2\rangle \right) \frac{\kappa_M^2 \omega_0^2}{4 m L^2} =\\
    & \nonumber = \frac{\omega_0 \omega_1}{4 m} \left(  \langle \hat\rho^2 \hat{p}_z\rangle + \langle \hat{p}_z \frac{\omega_1 \hat z}{\omega_0} \hat \rho^2 \rangle + \langle \hat\rho^2 \hat{p}_z\rangle + \langle \frac{\omega_1 \hat z}{\omega_0} \hat{p}_z \hat \rho^2 \rangle\right) \approx \\
    &  \approx \frac{\omega_0 \omega_1}{4 m} 2 \langle \hat\rho^2 \hat{p}_z\rangle^{(0)} \approx \frac{\omega_0 \omega_1}{2 m} \langle \hat\rho^2 \rangle^{(0)} \langle \hat{p}_z \rangle^{(0)}.
\end{align}

\section{RMS velocity for free-space Laguerre-Gaussian packet \label{sec:2mult}}
%================================================================================================
To evaluate RMS velocity \eqref{eq:velfree} we introduce the following functions
\begin{equation}
\begin{aligned}
    &Y_m=\int\limits_{0}^{\infty}y^m \left(\mathcal{L}_n^{|l|}(y)\right)^2\exp(-y)dy,\\
    &X_{m,k}=\int\limits_{0}^{\infty}y^m \mathcal{L}_n^{|l|}(y)\frac{\partial^k \mathcal{L}_n^{|l|}(y)}{\partial y^k}\exp(-y)dy=\\
    &(-1)^k\int\limits_{0}^{\infty}y^m \mathcal{L}_n^{|l|}(y)\mathcal{L}_{n-k}^{|l|+k}(y)\exp(-y)dy.
\end{aligned}
\end{equation}
RMS velocity can then be written as follows
\begin{equation}
\begin{aligned}
    &\expv{\hat{u}^2_{\perp}}=Y_l\left(1-\frac{it}{t_d}\right)N^2\frac{2\pi}{\sigma_{\perp}^2(t)}(l+1)-\\
    &Y_{l+1}N^2\frac{\pi}{\sigma_{\perp}(t)^2}\left(1-\frac{it}{t_d}\right)^2-X_{l,1}N^2\frac{4\pi (l+1)}{\sigma_{\perp}^2(t)}-\\
    &\frac{4\pi}{\sigma_{\perp}^2(t)}N^2X_{l+1,2}+\frac{4\pi\left(1-\frac{it}{t_d}\right)}{\sigma_{\perp}^2(t)}N^2X_{l-1,1}.
\end{aligned}
\end{equation}
$N$ is the normalization constant and evaluates to
\begin{equation}
    N^2 = \frac{1}{\pi l!}{n+l \choose n}^{-1}.
\end{equation}

$Y_m$ and $X_{m,k}$ functions can be expressed as coefficients in the series expansion of
\begin{equation}
    \begin{aligned}
        &Z_{\alpha,\beta,k}(s_1,s_2) = \sum\limits_{n,m}Z_{\alpha,\beta,k}\big|_{n,m}s_1^ns_2^m =\\
        &\int\limits_{0}^{\infty}U_{\alpha}(y,s_1)y^{l+k}U_{\beta}(y,s_2)\exp(-y)dy = \\
        &\frac{(1-s_1)^{l+k-\alpha}(1-s_2)^{l+k-\beta}}{(1-s_1s_2)^{l+k+1}}\Gamma(l+k+1),\\
    \end{aligned}
\end{equation}
where $U_{\alpha}(s,y)$ is the generating function for Laguerre polynomials with angular momentum $\alpha$ defined by \eqref{eq:gen}
\begin{equation}
    U_{\alpha}(s,y) = \sum\limits_{n=0}^{\infty}\mathcal{L}_n^{|l|}(y)s^n.
    \label{eq:gen}
\end{equation}
The connection between $Y_m$,$X_{m,k}$ and $Z_{\alpha,\beta,k}(s_1,s_2)$ is given by the following expressions
\begin{equation}
    \begin{aligned}
    &Y_l=Z_{l,l,0}\big|_{n,n}, Y_{l+1}=Z_{l,l,1}\big|_{n,n},\\ &X_{l-1,1} = -Z_{l,l+1,-1}\big|_{n,n-1},\\
    &X_{l,1}=-Z_{l,l+1,0}\big|_{n,n-1},X_{l+1,2}=Z_{l,l+2,1}\big|_{n,n-2}.
\end{aligned}
\end{equation}
Explicit expressions for $Y_m$ and $X_{m,k}$ read
\begin{equation}
    \begin{aligned}
    &Y_l=(-1)^nl!{-l-1 \choose n}=l!{n+l \choose n},\\
    &Y_{l+1} = (l+1)!\left[{n+l+1 \choose n}+{n+l \choose n-1}\right],\\
    &X_{l-1,1} = \sum\limits_{k=1}^{n}{-2 \choose k-1}{-l \choose n-k}=-\frac{(l+n)!}{(n-1)!},\\
    &X_{l,1} = X_{l+1,2} = 0.
    \end{aligned}
    \label{eq:YandX}
\end{equation}
With the help of \eqref{eq:YandX} the answer is obtained to be
\begin{align}
    \langle\hat{u}_{\perp}^2 \rangle_0=\frac{1}{m^2\sigma_r^2}\left(2n+|l|+1 \right).
\end{align}

\newpage
\bibliographystyle{unsrt}
\bibliography{references}

\begin{thebibliography}{10}

\bibitem{Allen1999}
L.~Allen, M.J. Padgett, and M.~Babiker.
\newblock {IV} the orbital angular momentum of light.
\newblock volume~39 of {\em Progress in Optics}, pages 291--372. Elsevier,
  1999.

\bibitem{FrankeArnold2008}
S.~Franke-Arnold, L.~Allen, and M.~Padgett.
\newblock Advances in optical angular momentum.
\newblock {\em Laser \& Photonics Reviews}, 2(4):299--313, 2008.

\bibitem{Bliokh2007}
Konstantin~Yu. Bliokh, Yury~P. Bliokh, Sergey Savel'ev, and Franco Nori.
\newblock Semiclassical dynamics of electron wave packet states with phase
  vortices.
\newblock {\em Phys. Rev. Lett.}, 99:190404, Nov 2007.

\bibitem{McMorran2011}
Benjamin~J. McMorran, Amit Agrawal, Ian~M. Anderson, Andrew~A. Herzing,
  Henri~J. Lezec, Jabez~J. McClelland, and John Unguris.
\newblock Electron vortex beams with high quanta of orbital angular momentum.
\newblock {\em Science}, 331(6014):192--195, 2011.

\bibitem{Tamburini2006}
F.~Tamburini, G.~Anzolin, G.~Umbriaco, A.~Bianchini, and C.~Barbieri.
\newblock Overcoming the rayleigh criterion limit with optical vortices.
\newblock {\em Phys. Rev. Lett.}, 97:163903, Oct 2006.

\bibitem{ct1}
Victor~V. Dodonov and Olga~V. Man'ko.
\newblock Universal invariants of quantum-mechanical and optical systems.
\newblock {\em J. Opt. Soc. Am. A}, 17(12):2403--2410, Dec 2000.

\bibitem{Mono}
Juan~P Torres and Lluis Torner.
\newblock {\em Twisted Photons: Applications of Light with Orbital Angular
  Momentum}.
\newblock John Wiley $\&$ Sons, Hoboken, NJ, 2011.

\bibitem{ct3}
{\em The Angular Momentum of Light}.
\newblock Cambridge University Press, 2012.

\bibitem{ct4}
K.Y. Bliokh, I.P. Ivanov, G.~Guzzinati, L.~Clark, R.~{Van Boxem},
  A.~B{\'e}ch{\'e}, R.~Juchtmans, M.A. Alonso, P.~Schattschneider, F.~Nori, and
  J.~Verbeeck.
\newblock Theory and applications of free-electron vortex states.
\newblock {\em Physics Reports}, 690:1--70, 2017.
\newblock Theory and applications of free-electron vortex states.

\bibitem{ct5}
V.~Serbo, I.~P. Ivanov, S.~Fritzsche, D.~Seipt, and A.~Surzhykov.
\newblock Scattering of twisted relativistic electrons by atoms.
\newblock {\em Phys. Rev. A}, 92:012705, Jul 2015.

\bibitem{ct6}
V.~A. Zaytsev, V.~G. Serbo, and V.~M. Shabaev.
\newblock Radiative recombination of twisted electrons with bare nuclei: Going
  beyond the born approximation.
\newblock {\em Phys. Rev. A}, 95:012702, Jan 2017.

\bibitem{ct7}
D.~V. Karlovets, G.~L. Kotkin, V.~G. Serbo, and A.~Surzhykov.
\newblock Scattering of twisted electron wave packets by atoms in the born
  approximation.
\newblock {\em Phys. Rev. A}, 95:032703, Mar 2017.

\bibitem{ct8}
Charles~W. Clark, Roman Barankov, Michael~G. Huber, Muhammad Arif, David~G.
  Cory, and Dmitry~A. Pushin.
\newblock Controlling neutron orbital angular momentum.
\newblock {\em Nature}, 525(7570):504--506, 2015.

\bibitem{ct9}
Alon Luski, Yair Segev, Rea David, Ora Bitton, Hila Nadler, A.~Ronny Barnea,
  Alexey Gorlach, Ori Cheshnovsky, Ido Kaminer, and Edvardas Narevicius.
\newblock Vortex beams of atoms and molecules.
\newblock {\em Science}, 373(6559):1105--1109, 2021.

\bibitem{FarEff}
Colin Greenshields, Robert~L Stamps, and Sonja Franke-Arnold.
\newblock Vacuum faraday effect for electrons.
\newblock {\em New Journal of Physics}, 14(10):103040, oct 2012.

\bibitem{Bliokh2012}
Konstantin~Y. Bliokh, Peter Schattschneider, Jo~Verbeeck, and Franco Nori.
\newblock Electron vortex beams in a magnetic field: A new twist on landau
  levels and aharonov-bohm states.
\newblock {\em Phys. Rev. X}, 2:041011, Nov 2012.

\bibitem{Silenko2021}
Liping Zou, Pengming Zhang, and Alexander~J. Silenko.
\newblock General quantum-mechanical solution for twisted electrons in a
  uniform magnetic field.
\newblock {\em Phys. Rev. A}, 103:L010201, Jan 2021.

\bibitem{Karlovets2021}
Dmitry Karlovets.
\newblock Vortex particles in axially symmetric fields and applications of the
  quantum busch theorem.
\newblock {\em New Journal of Physics}, 23(3):033048, mar 2021.

\bibitem{JAGA95}
S.~A. Khan and R.~Jagannathan.
\newblock Quantum mechanics of charged-particle beam transport through magnetic
  lenses.
\newblock {\em Phys. Rev. E}, 51:2510--2515, Mar 1995.

\bibitem{JAGA89}
R.~Jagannathan, R.~Simon, E.C.G. Sudarshan, and N.~Mukunda.
\newblock Quantum theory of magnetic electron lenses based on the dirac
  equation.
\newblock {\em Physics Letters A}, 134(8):457--464, 1989.

\bibitem{JAGA90}
R.~Jagannathan.
\newblock Quantum theory of electron lenses based on the dirac equation.
\newblock {\em Phys. Rev. A}, 42:6674--6689, Dec 1990.

\bibitem{NUF21}
Abhijeet Melkani and S.~J. van Enk.
\newblock Electron vortex beams in nonuniform magnetic fields.
\newblock {\em Phys. Rev. Research}, 3:033060, Jul 2021.

\bibitem{ST}
A.A. Sokolov and I.M. Ternov.
\newblock {\em Relativistic electron}.
\newblock Nauka, Moscow, 1974.

\bibitem{BLP}
V.B. Berestetskii, E.M. Lifshitz, and L.P. Pitaevskii.
\newblock {\em Quantum Electrodynamics}.
\newblock Oxford: Pergamon, 1982.

\bibitem{Peskin}
Michael~E. Peskin and Daniel~V. Schroeder.
\newblock {\em {An Introduction to quantum field theory}}.
\newblock Addison-Wesley, Reading, USA, 1995.

\bibitem{UFN}
B.A. Knyazev and V.G. Serbo.
\newblock Beams of photons with nonzero projections of orbital angular momenta:
  new results.
\newblock {\em Phys.-Usp.}, 61:449, 2018.

\bibitem{IvanovPubl}
Igor~P. Ivanov.
\newblock Promises and challenges of high-energy vortex states collisions.
\newblock {\em Progress in Particle and Nuclear Physics}, page 103987, 2022.

\bibitem{PDG}
M.~Tanabashi, K.~Hagiwara, K.~Hikasa, K.~Nakamura, Y.~Sumino, F.~Takahashi,
  J.~Tanaka, K.~Agashe, G.~Aielli, C.~Amsler, M.~Antonelli, D.~M. Asner,
  H.~Baer, Sw. Banerjee, R.~M. Barnett, T.~Basaglia, C.~W. Bauer, J.~J. Beatty,
  V.~I. Belousov, J.~Beringer, S.~Bethke, A.~Bettini, H.~Bichsel, O.~Biebel,
  K.~M. Black, E.~Blucher, O.~Buchmuller, V.~Burkert, M.~A. Bychkov, R.~N.
  Cahn, M.~Carena, A.~Ceccucci, A.~Cerri, D.~Chakraborty, M.-C. Chen, R.~S.
  Chivukula, G.~Cowan, O.~Dahl, G.~D'Ambrosio, T.~Damour, D.~de~Florian,
  A.~de~Gouv\^ea, T.~DeGrand, P.~de~Jong, G.~Dissertori, B.~A. Dobrescu,
  M.~D'Onofrio, M.~Doser, M.~Drees, H.~K. Dreiner, D.~A. Dwyer, P.~Eerola,
  S.~Eidelman, J.~Ellis, J.~Erler, V.~V. Ezhela, W.~Fetscher, B.~D. Fields,
  R.~Firestone, B.~Foster, A.~Freitas, H.~Gallagher, L.~Garren, H.-J. Gerber,
  G.~Gerbier, T.~Gershon, Y.~Gershtein, T.~Gherghetta, A.~A. Godizov,
  M.~Goodman, C.~Grab, A.~V. Gritsan, C.~Grojean, D.~E. Groom, M.~Gr\"unewald,
  A.~Gurtu, T.~Gutsche, H.~E. Haber, C.~Hanhart, S.~Hashimoto, Y.~Hayato, K.~G.
  Hayes, A.~Hebecker, S.~Heinemeyer, B.~Heltsley, J.~J. Hern\'andez-Rey,
  J.~Hisano, A.~H\"ocker, J.~Holder, A.~Holtkamp, T.~Hyodo, K.~D. Irwin, K.~F.
  Johnson, M.~Kado, M.~Karliner, U.~F. Katz, S.~R. Klein, E.~Klempt, R.~V.
  Kowalewski, F.~Krauss, M.~Kreps, B.~Krusche, Yu.~V. Kuyanov, Y.~Kwon,
  O.~Lahav, J.~Laiho, J.~Lesgourgues, A.~Liddle, Z.~Ligeti, C.-J. Lin,
  C.~Lippmann, T.~M. Liss, L.~Littenberg, K.~S. Lugovsky, S.~B. Lugovsky,
  A.~Lusiani, Y.~Makida, F.~Maltoni, T.~Mannel, A.~V. Manohar, W.~J. Marciano,
  A.~D. Martin, A.~Masoni, J.~Matthews, U.-G. Mei\ss{}ner, D.~Milstead, R.~E.
  Mitchell, K.~M\"onig, P.~Molaro, F.~Moortgat, M.~Moskovic, H.~Murayama,
  M.~Narain, P.~Nason, S.~Navas, M.~Neubert, P.~Nevski, Y.~Nir, K.~A. Olive,
  S.~Pagan~Griso, J.~Parsons, C.~Patrignani, J.~A. Peacock, M.~Pennington,
  S.~T. Petcov, V.~A. Petrov, E.~Pianori, A.~Piepke, A.~Pomarol, A.~Quadt,
  J.~Rademacker, G.~Raffelt, B.~N. Ratcliff, P.~Richardson, A.~Ringwald,
  S.~Roesler, S.~Rolli, A.~Romaniouk, L.~J. Rosenberg, J.~L. Rosner, G.~Rybka,
  R.~A. Ryutin, C.~T. Sachrajda, Y.~Sakai, G.~P. Salam, S.~Sarkar, F.~Sauli,
  O.~Schneider, K.~Scholberg, A.~J. Schwartz, D.~Scott, V.~Sharma, S.~R.
  Sharpe, T.~Shutt, M.~Silari, T.~Sj\"ostrand, P.~Skands, T.~Skwarnicki, J.~G.
  Smith, G.~F. Smoot, S.~Spanier, H.~Spieler, C.~Spiering, A.~Stahl, S.~L.
  Stone, T.~Sumiyoshi, M.~J. Syphers, K.~Terashi, J.~Terning, U.~Thoma, R.~S.
  Thorne, L.~Tiator, M.~Titov, N.~P. Tkachenko, N.~A. T\"ornqvist, D.~R. Tovey,
  G.~Valencia, R.~Van~de Water, N.~Varelas, G.~Venanzoni, L.~Verde, M.~G.
  Vincter, P.~Vogel, A.~Vogt, S.~P. Wakely, W.~Walkowiak, C.~W. Walter,
  D.~Wands, D.~R. Ward, M.~O. Wascko, G.~Weiglein, D.~H. Weinberg, E.~J.
  Weinberg, M.~White, L.~R. Wiencke, S.~Willocq, C.~G. Wohl, J.~Womersley,
  C.~L. Woody, R.~L. Workman, W.-M. Yao, G.~P. Zeller, O.~V. Zenin, R.-Y. Zhu,
  S.-L. Zhu, F.~Zimmermann, P.~A. Zyla, J.~Anderson, L.~Fuller, V.~S. Lugovsky,
  and P.~Schaffner.
\newblock Review of particle physics.
\newblock {\em Phys. Rev. D}, 98:030001, Aug 2018.

\bibitem{Reiser}
M.~Reiser.
\newblock {\em Theory and Design of Charged Particle Beams}.
\newblock Wiley, New York, 2008.

\bibitem{Cho}
B.~Cho, T.~Ichimura, R.~Shimizu, and C.~Oshima.
\newblock Quantitative evaluation of spatial coherence of the electron beam
  from low temperature field emitters.
\newblock {\em Phys. Rev. Lett.}, 92:246103, Jun 2004.

\bibitem{Cho2013}
B.~Cho and C.~Oshima.
\newblock Electron beam coherency determined from interferograms of carbon
  nanotubes.
\newblock {\em Bulletin of the Korean Chemical Society}, 34:892–898, 2013.

\bibitem{Lat}
Tatiana Latychevskaia.
\newblock Spatial coherence of electron beams from field emitters and its
  effect on the resolution of imaged objects.
\newblock {\em Ultramicroscopy}, 175:121--129, 2017.

\bibitem{Ehberger}
Dominik Ehberger, Jakob Hammer, Max Eisele, Michael Kr\"uger, Jonathan Noe,
  Alexander H\"ogele, and Peter Hommelhoff.
\newblock Highly coherent electron beam from a laser-triggered tungsten needle
  tip.
\newblock {\em Phys. Rev. Lett.}, 114:227601, Jun 2015.

\bibitem{PRA19}
Dmitry Karlovets.
\newblock Dynamical enhancement of nonparaxial effects in the electromagnetic
  field of a vortex electron.
\newblock {\em Phys. Rev. A}, 99:043824, Apr 2019.

\bibitem{Landau}
L.~D. Landau and E.~M. Lifshitz.
\newblock {\em Quantum Mechanics: Nonrelativistic Theory}.
\newblock Butterworth-Heinemann, Burlington, Massachusetts, 1981.

\bibitem{N2opt}
M.~Newstein and B.~Rudman.
\newblock Laguerre-gaussian periodically focusing beams in a quadratic index
  medium.
\newblock {\em IEEE Journal of Quantum Electronics}, 23(5):481--482, 1987.

\bibitem{Karlovets_paraxial1}
Dmitry Karlovets.
\newblock Relativistic vortex electrons: Paraxial versus nonparaxial regimes.
\newblock {\em Phys. Rev. A}, 98:012137, Jul 2018.

\bibitem{TESLA}
B.~Aune, R.~Bandelmann, D.~Bloess, B.~Bonin, A.~Bosotti, M.~Champion,
  C.~Crawford, G.~Deppe, B.~Dwersteg, D.~A. Edwards, H.~T. Edwards,
  M.~Ferrario, M.~Fouaidy, P.-D. Gall, A.~Gamp, A.~G\"ossel, J.~Graber,
  D.~Hubert, M.~H\"uning, M.~Juillard, T.~Junquera, H.~Kaiser, G.~Kreps,
  M.~Kuchnir, R.~Lange, M.~Leenen, M.~Liepe, L.~Lilje, A.~Matheisen, W.-D.
  M\"oller, A.~Mosnier, H.~Padamsee, C.~Pagani, M.~Pekeler, H.-B. Peters,
  O.~Peters, D.~Proch, K.~Rehlich, D.~Reschke, H.~Safa, T.~Schilcher,
  P.~Schm\"user, J.~Sekutowicz, S.~Simrock, W.~Singer, M.~Tigner, D.~Trines,
  K.~Twarowski, G.~Weichert, J.~Weisend, J.~Wojtkiewicz, S.~Wolff, and
  K.~Zapfe.
\newblock Superconducting tesla cavities.
\newblock {\em Phys. Rev. ST Accel. Beams}, 3:092001, Sep 2000.

\end{thebibliography}
\end{document}